# "Evaluation and Implementation of Machine Learning Algorithms to Predict Early Detection of Kidney and Heart Disease in Diabetic Patients"

**SYED IBAD HASNAIN**
*Biomedical Engineering*
*Sir Syed University of Engineering and Technology*
*May 2025*

# TABLE OF CONTENTS

# LIST OF FIGURES

# LIST OF TABLES

# LIST OF ABBREVIATIONS

| | |
|---|---|
| **CKD** | Chronic Kidney Disease |
| **CVD** | Cardiovascular Disease |
| **ML** | Machine Learning |
| **AUC** | Area under Curve |
| **AI** | Artificial Intelligence |
| **DM** | Diabetes Mellitus |
| **HbA1c** | Hemoglobin A1c (glycated hemoglobin) |
| **LR** | Logistic Regression |
| **MI** | Myocardial Infarction |
| **RF** | Random Forest |
| **ROC** | Receiver Operating Characteristic (curve) |
| **SMOTE** | Synthetic Minority Over-sampling Technique |
| **SVM** | Support Vector Machine |
| **CV** | Cross Validation |

# LIST OF NOTATIONS

| | |
|---|---|
| $\sigma$ | Standard Deviation |
| $\sigma^2$ | Variance |
| $\mu$ | Mean |
| $Z$ | Z-score (standard normal deviate) |
| $d$ | Margin of error (usually 0.05) |
| $C$ | Regularization parameter in SVM |
| $F$ | F-statistic in ANOVA |
| $k$ | Number of cross-validation folds |
| $n$ | Sample size |

# ABSTRACT

Cardiovascular disease (CVD) and chronic kidney disease (CKD) are important complications of diabetes, resulting in high morbidity and mortality. The early diagnosis of these conditions is vital for appropriate treatment, but traditional marker-based diagnostics tend to lack sensitivity in the early phases. This research combines conventional statistical methods with advanced machine learning approaches to enhance the early diagnosis of CKD and CVD among diabetic patients.

Descriptive and inferential statistical procedures were first computed and examined through statistical software SPSS to assess CVD and CKD relations to clinical and demographic parameters. Patients were grouped on basis of their disease status i.e Group A (Patients with both Kidney and heart disease), Group B ( Patients with only kidney disease), Group C (Patients with only heart disease) and Group D (Patients with no disease). Descriptive analysis showed clear patterns of poor health indicators with Group A, Group B and Group C. Notably, Serum Creatinine, Hypertension showed strong correlation in CKD group whereas Cholesterol, Triglycerides, History of Myocardial Infarction, Stroke and Hypertension showed strong correlation in CVD group. Statistical results revealed that patients with CKD and CVD has a substantial impact on key clinical parameters and guided in feature selection for machine learning models.

Machine learning models including Logistic regression, Support Vector Machine and Random Forest were implemented. These models showed promising results, especially in identifying patients at risk of CKD.

In conclusion, ensemble models, particularly Random Forest, demonstrated great capability for high-risk diabetic patient identification, especially for CKD. SPSS analysis validated the statistical importance of core features, validating their integration into the machine learning pipeline. There are still hurdles to overcome such as interpretability and class imbalance; however, these models provide substantial advancement over conventional diagnostic methods.

# Chapter 1
# INTRODUCTION

In recent years, the science of medicine has progressed in relevance correlatively with the advent of technological innovations. One of the prominent advancements is in the area of diagnostic procedures, where powerful ML (machine learning) algorithms can be employed for predictive analysis, disease diagnosis, and even clinical decision support systems[1]. Diabetes mellitus is among those chronic conditions that pose a major threat for public health because, within the scope of a single pathology, it includes a large amount of fatalities and serves as a precursor of severe associated comorbidities like diabetic nephropathy, a chronic progressive illness of the kidneys which leads to kidney failure augmentation, and diabetic cardiomyopathy, a condition in which heart muscle tissues gradually get replaced with scar tissues[2]. The progression of these diseases reduces people's lives, speeds up mortality increase, and makes healthcare much more expensive. Skeletal complications are some of the most prevalent forms of diabetes related complications and, if the disease is not identified in time, can lead to amputation [3].

Most diabetic nephropathy and cardiomyopathy cases are symptomless during the initial phases, making it difficult to diagnose them without routine checkups [3]. Blood tests and imaging, like other diagnostic techniques, are expensive, and their execution takes time. Furthermore, they often fail to identify diseases in their early, treatable stages. These factors further postpone detection and intervention, resulting in organ damage that is often irreversible. All of these issues demonstrate the necessity for more effective tools for diagnosis. The use of machine learning algorithms in practical medicine might resolve this challenge through reliable, expedited, and economical approaches for early disease risk prognosis[4].

Projected and automated systems procedures through Diabetic AI Care is built around machine learning, which is a subset of artificial intelligence (AI). It allows devices to learn from the data automatically and make a decision or prediction without manual algorithms integration[5]. Automated analysis allows to recognize clusters within huge open-ended reports and sets of information that include gender, age, medical background, diagnostic examinations, and personal lifestyle, which might signify the emergence of health issues. This technology also accelerates and enhances accuracy in the diagnostic processes by identifying patterns that are often too delicate for human practitioners to notice, consequently shifting disease management to a prevention-centered strategy [6].

This study attempts to address the unmet need for developing efficient screening methods for kidney and heart complications among diabetic patients [7]. To accomplish this, it is expected that upon the application of machine learning algorithms on clinical data, predictive models will be created that will accurately estimate the probability of disease occurrence for high-risk patients [6]. This not only improves the chances of identifying the problem earlier, but ensures action can be undertaken sooner that may stop the disease from advancing to a more debilitating stage later on.

## 1.1 Motivation

Motivation is simpler than a statement; it is a driving force originating from pure practical problems. The need for better diagnostic tools has never been more critical in the context of healthcare, especially in the care of diabetes and its complications. Diabetes is a chronic disease that is often detected too late, because its beginning usually runs without any symptoms. Most often the complications of diabetes like kidney and heart diseases are only detected when they advance to a stage of irreversible damage. The reason I became inspired to do this research is because there is a gap in early detection.

However, traditionally, kidney and heart diseases – especially in diabetic patients – are not easily diagnosed and monitored at their earliest time due to the fact that these methods are not up to date. The signs can be subtle and may not appear in a form that can be acted upon in the short term. The critical issue is that these diseases are detectable only in the late stages of the disease and therefore, early intervention is the key to prevent the progression of these diseases. The goal we want to accomplish in this study is to investigate how the tool of machine learning, which is a powerful means to analyze vast and bulky datasets, can be utilized in predicting the onset of kidney and heart diseases in diabetic patients in the early stage. Moreover, machine learning algorithms can use patient data to recognize patterns and trends that are otherwise invisible to human eyes and offer the ability to earlier and more accurate determine patient diagnosis to healthcare professionals.

This represents a fundamental shift toward proactive healthcare in biomedical engineering. Now, it's not about just analyze data, it's about understanding data and being able to apply it. These insights can be used by healthcare providers to discover high risk patients, introduce counter action earlier and all together improve outcomes. It is not an intent to replace clinicians, but rather to supply them with another device to help them make their decisions and become more precise and efficient at the same time. Take the case of monitoring conditions like arthritis when we follow specific symptoms (swelling or limited mobility) to determine its severity. Predicting kidney and heart disease in diabetic patients could be managed by machine learning in a similar way. These algorithms analyze data points like medical history, lab results and lifestyle factors to help healthcare providers identify people who are at risk for certain conditions well before clinical symptoms appear over physicians' heads. Detection early is important because that allows doctors and patients time to act before the disease has progressed to a point when treatment options are less effective and less affordable.
Machine learning makes it possible to combine diagnosis in a simple way. Additionally, it can also be used to optimize patient care by improving the accuracy of risk assessments and rapidly intervening. By taking a proactive approach, the quality of life of patients will also improve, saving healthcare systems from the expenses of late stage diseases and costly intensive treatments.

Therefore, the motivation behind this research is to fill the gap in diabetic early disease detection. This study shows how machine learning can make healthcare more predictive, more personal, and ultimately better. The ultimate purpose is to supply healthcare providers with more powerful tools and patients with more power to manage their health, and in the long run, this reduces the lasting impacts of kidney and heart diseases. Early detection of the disorder and timely intervention offers the opportunity for significant improvement in diabetic patients' outcomes, and improves the efficiency of the entire healthcare system.

## 1.2 Thesis contribution

This research's main objective is to create, assess, and implement machine learning models for the early detection of kidney and heart diseases in patients suffering from diabetes. This study attempts to apply advanced computation techniques to solve problems in real life healthcare. The goal is to develop predictive models that can detect individuals with a high probability of developing kidney and heart diseases years before clinical diagnosis is made so that healthcare providers can intervene sooner, allowing for more effective management of the diseases.

The specific contributions of this thesis are as follows:

i. It analyzes a number of machine learning techniques and their predictive capabilities with respect to kidney and heart diseases in diabetic patients. The study will apply decision trees, random forests, support vector machines (SVM), and neural networks and evaluate each based on accuracy, sensitivity, specificity, and other relevant factors related to disease prediction.

ii. The feature selection and data preprocessing steps are analyzed to determine their impacts on the performance of the machine learning models. Medical data is often disordered and lacking, so it is extremely important to clean, preprocess, and extract relevant features for the model training. The study aims to determine the accuracy of the models with different feature selection techniques of correlation analysis and principal component analysis (PCA).

iii. This dissertation will prove the possibility of applying automated learning algorithms into real-life clinical practices. The purpose is to proactively provide the predictive tools for the personnel that will enable them to detect potential diabetic patients at high risk of developing kidney or heart failure diseases. This

   shift has the potential to increase patient management, better healthcare results, and lower costs associated with healthcare.
iv. Different machine learning models will be analyzed in regard to their performance using specified comparative evaluation metrics. The models' capabilities in predicting the early signs of kidney and heart diseases will be measured using accuracy, precision, recall, F1-score, and area under the curve (AUC) performance indicators.

This research seeks to reduce the burden of chronic disease on patients and health systems by improving the early diagnosis and management of heart and kidney complications in diabetics. This objective will be accomplished by developing reliable, data-based tools for disease forecasting.

## 1.3 Thesis Structure

Chapter 1 provides an introduction to the study and a brief state of the artificial intelligence in diagnosis of medical needs implement in health care system as well as thesis motivation and its contribution in field of biomedical engineering.

Chapter 2 provides detailed background knowledge for the study overview of diabetes mellitus, Chronic kidney disease, Cardiovascular disease, biomarkers, comorbidities observed among diabetic patients, role of machine learning in medical treatment of diabetic patients with CVD and CKD.

Chapter 3 describes about the study of algorithm of Machine learning. This proposed study is established using the machine learning algorithm generated from physical and clinical dataset of the diabetic patients. These machine learning algorithm are associated with accuracy and evaluation matrices.

Chapter 4 provide results from the previous chapters to demonstrate the accuracy and evaluation matrices by the proposed machine learning algorithms.

Chapter 5 concluded with a summary of the key findings from the research conducted here. There is also a discussion of future work to be carried out in the area of machine learning for diabetic patients to better decision making tools for patients with chronic conditions.

# Chapter 2
# LITERATURE REVIEW

## 2.1 Diabetes mellitus

Diabetes mellitus (DM) is a chronic metabolic disorder that is affecting the whole world today. Roughly 537 million adults were living with diabetes in the year 2021 alone, and it is estimated that if this trend continues, that number will reach 643 million by 2030 and 783 million by 2045[8]. Diabetes can be separated into two categories; Type 1 diabetes (T1D) and Type 2 diabetes (T2D), the latter of which is the predominating type making up over 90% of all cases. T1D on the other hand is an autoimmune disorder that hinders beta, the insulin-producing cells in the pancreas, resulting in a body that lacks insulin[9].

The difficulties that accompany diabetes are changeable and tremendously damaging. Cardiovascular disease (CVD) remains among the most prevalent and critical issues, where diabetes increases the risk of CVD by two to four times[10]. Hyperglycemia, dyslipidemia, and hypertension aid in the processes associated with endothelial dysfunction, inflammation, and atherosclerosis, all of which advance CVD. In the same fashion, diabetic kidney disease (DKD) is one of the leading etiologies of end stage renal disease (ESRD), affecting nearly 40% of diabetic individuals [11]. The pathophysiology of DKD is characterized by hyperglycemic glomerular hyperfiltration, oxidative stress, and chronic inflammation, which gradually constrain the nephron's functional capacity[12]. Recent research revealed a significant increase in the prevalence of CKD in people suffering from diabetes, where, distinctly, one in three adults with diabetes exhibiting some form of renal impairment [3]. This disturbing reality stresses the need for early intervention to avert the progression to ESRD.

Besides macrovascular problems such as cardiovascular disease, diabetes also causes microvascular ones like neuropathy and retinopathy [13]. Diabetic neuropathy occurs in up to 50 percent of persons with diabetes and presents as pain, numbness, and increased risk for foot ulcers and amputations. Retinopathy, in diabetic patients, is a leading cause of blindness in adults alongside hyperglycemia and damage to the blood vessels. According to a 2023 study by Cheung et al., one in three people with diabetes have retinopathy which shows the importance of proactive screening [14]. The healthcare expenditure for diabetes globally is equally shocking, estimated at about $966 billion in 2021 with expectations soaring in coming years.

Considering the increase in the number of diabetes cases diagnosed and the profound effects it has on people's health, there is more focus on using cutting edge technologies like machine learning (ML) for detection and prediction purposes. Random forest algorithms, support vector machines (SVM), and neural networks are very promising for pattern recognition and predicting results from large datasets. Recent article reviews have shown the effectiveness of ML in assessing risks for CVD and CKD complications in diabetes patients [15]. For example, Researchers were able to predict the 10-year risk of cardiovascular disease in diabetic patients with more than 85 percent accuracy, while Zhang et al. reported the development of an ML algorithm that comprises more than 90 percent sensitivity for CKD progression prediction among diabetics [16]. These advances in technology make it clear that effective ML implementation can help with the caring of patients with diabetes and ultimately reduce the complications. Seeing that the burden of diabetes is continuously increasing globally, it is vital to combine new innovations with traditional public health approaches to successfully tackle this important issue[17].

The ongoing diabetes pandemic has been alarming for many countries, and Pakistan is no exception. The newest IDF estimates suggest that around 33 million adults in Pakistan are diabetic, resulting in a national prevalence estimate of 26.7% – one of the highest in the world[18]. The rapid and unchecked increase in the prevalence of diabetes in Pakistan has been linked to the worsening health consequences of accelerated urbanization, widespread physical inactivity, unhealthy eating patterns, and the South Asian tendency towards genetic insulin resistance[19]. Besides, there is a glaring and compounding lack of healthcare access, low diabetes health literacy, and poor diabetes care. Recent studies provided insights in 2023 that nearly 50% of patients with diabetes in Pakistan are undiagnosed, leading to worse prognosis of them suffering from cardiovascular disease, kidney failure, and retinopathy [20]. Not to mention, Pakistan suffers from a colossal economic burden owing to unchecked hyperglycemia and noncompliance, as expenditures in diabetes treatment eat up a desirable part of an insufficient healthcare budget[21].

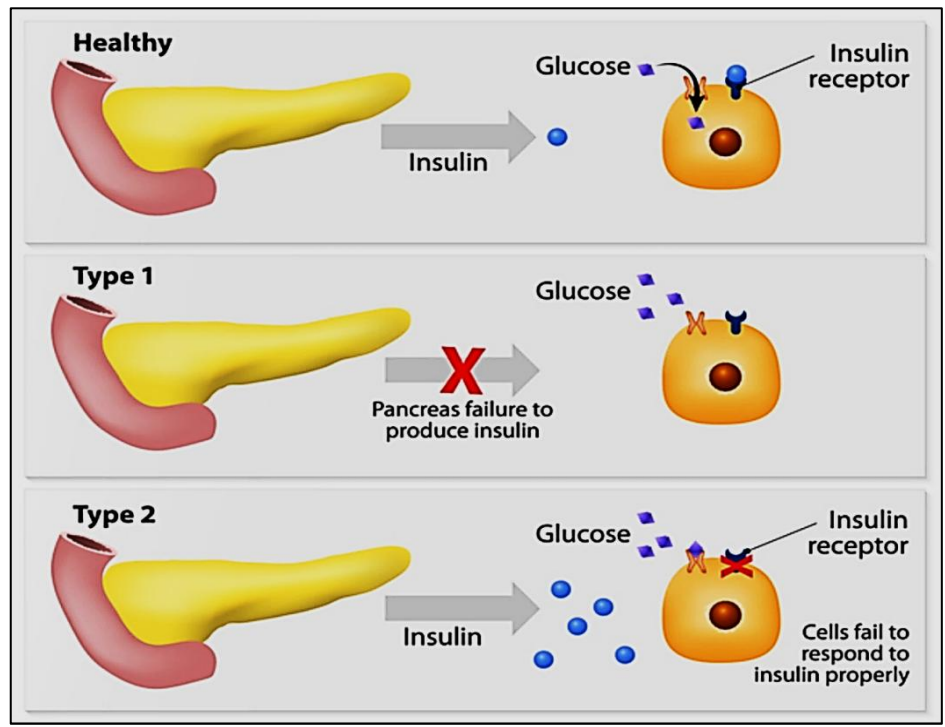

*Figure 2.1 Diabetes Type 1 and 2 schematic*

*Source: "Diabetes and Lactose intolerance and Correlation between Men and Women of Different Age Groups" - Scientific Figure on Research Gate. [Accessed 30 Apr 2025]*

The increasing rate of diabetes in Pakistan highlights the importance of tailored public health approaches with universal screening, patient education, and implementation of machine learning based early detection and anthropometric monitoring systems. If no action is taken, it is highly probable that the diabetes epidemic in Pakistan will deepen, which in turn will have negative impacts on the economy and on public health in the country[2, 18].

## 2.2 Epidemiology

Diabetes Mellitus is a growing global challenge. Its prevalence keeps increasing at an alarming speed. An estimate of 537 million adults were living with diabetes in 2021[2]. This figure will further escalate to 643 million in 2030 and 783 million by 2045. Diabetes is fueled by urbanization, sedentary lifestyles, unhealthy diets, and aging populations. 90 to 95 percent of diabetic people have Type 2 diabetes. The remaining percentage suffer from Type 1 and other forms of diabetes including gestational diabetes[22]. The currently reported prevalence of diabetes in adults aged 20-79 is 10.5 percent globally. The western pacific, South East Asia, and Middle East & North Africa regions are at the top in terms of reported diabetes cases[23].

The consequences of diabetes in low and middle income countries (LMICs) is significantly high when compared to their higher income counterparts. This difference can be attributed to unawareness, neglect, and difficulty in diabetes care customarily[24]. With this being said, the IDF stated in 2023 that 'approximately 33 million people' (26.7% of adults) in Pakistan are living with diabetes which makes the country one of the most diabetes prevalent regions in the world [2, 8]. India is reported to have over 74 million adults suffering from diabetes, making it the second highest country after China[25]. The US, a high-income nation, has a diabetes prevalence of 11.3% (about 37 million adults)[26]. Despite having lower prevalence rates, diabetes care remains unequal, as marginalized and ethnic community members remain at greater risk of diabetes and other complications [3, 22].

The economic strain presented by diabetes is enormous, with Global healthcare spending relating to diabetes costing $966 billion in 2021, which was approximately 9% of worldwide healthcare expenditure[27]. That number is likely to grow with growing diabetes cases. Diabetes remains a significant contributor to illness and death with 6.7 million deaths in 2021, translating to one death every five seconds. The disease is also a prime contributor to cardiovascular disease (CVD), chronic kidney disease (CKD), and various other diabetic complications which makes matters worse for the healthcare systems[28]. For example, recent studies show that 32.2% of people with

T2D have some form of CVD, and 40% of patients with diabetes go on to develop CKD[7].

The profile of diabetes calls for immediate action in the form of integrated public health approaches such as promotion, screening as well as timely and effective treatment[29].Efforts to stem the tide of diabetes must focus on alterable risk factors like obesity, sedentary lifestyle, and poor nutritional intake[30]. Also, the deployment of new technologies, such as machine learning for predictive analytics and tailored care delivery, can greatly impact the burden of diabetes and its complications worldwide. If no urgent and cohesive steps are taken, the diabetes epidemic will worsen and become an even greater danger to a country's health and welfare[31].

## 2.3 Comorbidities associated with DM

Diabetes mellitus (DM) is best characterized as a multifaceted metabolic disorder of unknown origin. It is often accompanied by a multitude of additional pathologies, which exacerbate the risk for morbidity and mortality. These additional pathologies may be grouped into macro- and microvascular complications, as well as multifactorial entities such as obesity and hypertension [32]. The existence of these additional pathologies contributes to the complexity of managing diabetes and highlights the importance of comprehensive care of the patient.

### 2.3.1 Cardiovascular Disease (CVD)

CVD is undoubtedly the most common and serious associated comorbidity that diabetes patients suffer with. Complications from uncontrolled diabetes put a person at two to four times greater risk of developing CVD compared to a non-diabetic individual[33]. This encompasses coronary artery disease (CAD), heart failure, stroke, and peripheral arterial disease (PAD). Some of the main contributors to the complications from CVD includes hyperglycemia, insulin resistance, dyslipidemia, chronic inflammation, which are now marked with endothelial dysfunction and atherosclerosis, which advance the progression of CVD[34]. Data published by Einarson et al in 2022 suggest that 32.2% of individuals diagnosed with Type 2 diabetes (T2D) also have some form of CVD, which has emerged as the commonest cause of death in diabetic patients[35].

**2.3.2 Chronic Kidney Disease (CKD)**

Diabetes is a major microvascular complication resulting diabetic kidney disease (DKD), which is also a leading cause of end-stage renal disease (ESRD). One research reported that roughly 40% of patients suffering from diabetes are likely to develop CKD[36]. The pathophysiology is involved with hyperglycemia, oxidative stress, and chronic inflammation resulting in glomerular hyper filtration which damages the kidneys over time. Since the early stages of CKD are usually asymptomatic, it is crucial to periodically check for albuminuria and eGFR in the diabetic population [7, 37].

**2.3.3 Diabetic Retinopathy**

Diabetic retinopathy is a greater concern for vision loss globally as it is one of the leading causes of blindness for elderly adults. It is estimated that about a third of diabetics suffer from retinopathy. Chronic hyperglycemia results in the damage of the blood vessels, impairing vision and can even result in blindness[33]. Regular eye exams and tight glycemic control is crucial in early detection to ensure the disease does progress.

**2.3.4 Diabetic Neuropathy**

Diabetic Neuropathy is a microvascular complication affecting up to 50% of those with diabetes. This condition manifests itself as Peripheral Neuropathy which consists of numbness, tingling, and pain in the extremities, as well as Autonomic Neuropathy, which affects the digestive system, heart, and bladder. In combination with other forms of Peripheral Neuropathy, these can significantly increase the risk of foot ulcers and amputations, especially in uncontrolled diabetic patients suffering from PAOD (Peripheral Arterial Occlusive Disease)[5].

**2.3.5 Hypertension**

As per the American Diabetes Association, it is estimated that 60-80% of diabetics are diagnosed with hypertension. Then having diabetes in combination with hypertension leads to a greater probability of CVD (Cardiovascular Disease) and CKD (Chronic Kidney Disease); due to the shared Insulin Resistance and Endothelial dysfunction along with an increase in the Renin Angiotensin Aldosterone system (RAAS)activation[38].

**2.3.6 Obesity and Metabolic Syndrome**

Type 2 diabetes, or T2D, has shown an alarming increase in the past decades. This puts the majority of patients suffering from Obesity, which can stem from lack of physical activity paired by insulin resistance, lipid abnormality and chronic inflammation that worsens the patients' risk of developing CVD as well as other associated heart complications or diseases[39]. Metabolic syndrome entails Abdominal obesity, Hypertension, Dyslipidemia along with hyperglycemia, resulting in increased cardiovascular deteriorating effects in diabetic patients[40].

**2.3.7 Non-Alcoholic Fatty Liver Disease (NAFLD)**

NAFLD is now being identified as a comorbidity of Diabetes Mellitus Type II, with estimates between 50-70% of individuals who suffer from T2D also having NAFLD. It can vary from simple steatosis to non-alcoholic steatohepatitis (NASH), which can progress towards cirrhosis as well as hepatocellular carcinoma [41]. Both insulin resistance and hyperglycemia are core features of the pathology and the progression of NAFLD[42].

**2.3.8 Mental Health Disorders**

With diabetes, comes the increased risk of suffering from other mental related issues such as depression, anxiety or even diabetes distress. Patients suffering from diabetes are more likely to be depressed on a scale which is two or even thrice that of the general population. The interplay between mental health and diabetes is complex and ultimately leads to poor resolution of diabetes self-management and an increase in glycemic control[43].

### 2.3.9 Infections

Patients who have diabetes are at a higher risk of infection as a result of both immune dysfunction and hyperglycemia, which forms a conducive setting for the pathogens to thrive[12]. Diabetes patients are more vulnerable to skin and soft tissue infections, and UTIs along with oral infections. Poorly managed diabetic foot infections are severe and can lead to amputations if control is not gained quickly.

### 2.3.10 Cancer

New evidence indicates a connection between diabetes and an increased risk for several types of cancer, including liver, pancreatic, colorectal, and breast cancer. It is postulated that chronic hyperglycemia, inflammation, and insulin resistance contributes towards the development of cancer. Individuals with diabetes are identified to have a 20% and 30% higher risk of developing cancer than people without diabetes, based on a 2023 research [1].

The various chronic comorbidities associated with diabetes are complex and interrelated which decidedly affects the patient's quality of life and health prognosis[44]. Managing diabetes effectively calls for a more holistic approach that bears in mind not only control of glycemic levels but also the risk of such comorbidities and their management. To mitigate diabetes complications and enhance outcomes, routine checks, proactive measures, and coordinated efforts by various healthcare providers is key[45].

## 2.4 Cardio vascular disease

Cardiovascular disease (CVD) is the most frequent and severe complication associated with diabetes mellitus, representing a notable portion of morbidity and mortality among diabetic patients. Diabetic individuals are two to four times more likely to have CVD than nondiabetic individuals[33]. CVD accounts for roughly 50-70% of deaths in these individuals. The combination of hyperglycemia, insulin resistance, dyslipidemia, and chronic inflammation brings about a pathological state for atherosclerosis development, as well as endothelial dysfunction and other heart health issues. The main cardiovascular conditions associated with diabetes are coronary artery disease (CAD), heart failure, stroke, and peripheral arterial disease (PAD)[46].

Diabetics having blockages in plaque-filled coronary arteries are certain to have their heart muscles fail as the blood supply to the heart is narrowed - a condition known as coronary artery disease. Hyperglycemia causes the formation of advanced glycation end products (AGEs), which in turn accelerate the stiffening of blood vessels as well as the instability of plaques [7]. A significant proportion of patients with heart failure, specifically those with preserved ejection fraction (HFpEF) and reduced ejection fraction (HFrEF), are diabetic - a condition known for its association with insulin resistance[47]. The dysfunctional cardiac conditions are the results of impaired metabolism of the heart muscles, accompanied by accumulation of fibrous tissue. People with diabetes are more prone to suffer from strokes - Ischemic strokes in particular – due to clot formations or, atherosclerosis of the arteries located in the brain. Within diabetes, lesions and self osteolytic infections of bones accompanied by inflammation and pus are often seen, but these are not the only changes[48].

CVD in diabetes has complex pathophysiology that can be interpreted in different ways. Among these mechanisms, chronic hyperglycemia with oxidative stress, chronic inflammation, and vascular damage atherogenesis is very common. Other important features in this scenario are the abnormal lipids: increased triglyceride and small dense LDL particles along with decreased HDL cholesterol are clear ingredients of higher cardiovascular distress. The presence of hypertension in 60-80% of diabetics is another addition to the burden of the diseased heart as it increases cardiac output, which along with the existing damage to blood vessels adds to the deteriorated health condition of the heart. Other than these, increased risk of diabetes-related acute cardiovascular events comes from diabetes-associated hypercoagulable states with disturbed platelet functions and other vascular pathologies.

To prevent complications further, one must always act sooner in the pre-clinical stages of CVD in diabetes. More clinically available markers associated with inflammation like high-sensitivity Troponin (hs-TnT), B-type Natriuretic Peptide (BNP) and C-reactive protein (CRP) along with more defined biomarker, added with conventional risk factors, allow for the estimation of CVD severity and monitoring its course. Non-pharmacological measures to control the weight, physical exercise, and healthy nutrition with any heart friendly diet are central in the primordial prevention of the disease. Statins, ACE inhibitors, SGLT2, and GLP-1 receptor agonists are some newly developed drugs that have marked astonishing superiority in dealing with diabetic patients in heart failure to prevent heart attack and or cardiovascular mortality [49]. E.g. the use of SGLT2 inhibitors decreases the odds of hospitalization due to heart failure by 30-35%, while the use of GLP-1 receptor agonist reduces the change of occurrence of major adverse cardiovascular events (MACE) by 12-26%.

To summarize, diabetes has many complications and one of the most common is cardiovascular disease, which is caused by a number of metabolic and vascular functions. Efficient treatment requires tailoring interventions for early diagnosis, lifestyle changes, and specific medication treatment. Targeting cardiovascular risk factors and utilizing innovative interventions can greatly alleviate the burden of CVD in diabetic individuals, thereby enhancing quality of life and life expectancy.

## 2.5 Chronic Kidney Disease

Chronic kidney disease (CKD) is potentially one of the most devastating complications of diabetes mellitus, impacting about 40% of diabetic patients and being the most common reason for end stage renal disease (ESRD) globally. Like type 1 and type 2 diabetes, CKD too has a plethora of risk factors which include but are not limited to chronic hyperglycemia, hypertension and other metabolic disorders. The DKD pathophysiology has multiple mechanisms, such as hyperglycemic induced glomerular hyper filtration, oxidative stress, chronic inflammation and Advanced Glycation End products (AGEs) phenomenal structural and functional harm to the kidney accumulate over time [7].

The progression of DKD and CKD is usually painless in the beginning, hence it is important to keep track of your health on a regular basis. Important diagnostic signs consist of the UACR test, which shows microalbuminuria (30-300 mg/g) or macro albuminuria (>300 mg/g) and eGFR. To measure kidney performance the eGFR is set at <60 mL/min/1.73 m² for three months or greater. This, without necessary action, can progress through 5 stages of DKD; in turn causing ESRD, and in need of dialysis or a kidney replacement [37].

The anatomical and physiological features of Diabetic Kidney Disease (DKD) have been dissected to better understand the medical implications of its risk factors which incorporate suboptimal glycemic control, hypertension, dyslipidemia, obesity, as well as smoking. Additionally, economic and sociological conditions impact the development of the disease[29]. Optimal management of DKD aims for slowing the deteriorating kidney function, while also preventing the cardiovascular complications that are very common among patients with diabetes and Chronic Kidney Disease (CKD). Appropriate strategies include intensive care in the regulation of glycemia with an HbA1c goal of <7% in the vast majority, and control of blood pressure with a target of <130/80 mmHg[50]. Pharmacological treatment with ACE inhibitors or angiotensin II receptor blockers (ARBs) is the first choice in the treatment of albuminuria, as well as in prevention of further renal deterioration.

Recently introduced SGLT2 inhibitors and GLP-1 receptor agonists have provided astounding changes regarding their renal protective properties. SGLT2 inhibitors such as Empagliflozin and Dapagliflozin are able to decrease the progression of kidney disease by 30-40%, in addition to reducing end-stage renal disease (ESRD) and subsequent cardiovascular complications[51]. Furthermore, GLP-1 receptor agonists, like liraglutide and semaglutide, offer robust renal protective effects, especially noted for their ability to decrease albuminuria. In addition, mineralocorticoid receptor antagonists (MRAs) like Finerenone are emerging as potent candidates to improve kidney and cardiovascular outcomes in diabetics with CKD.

Changes to one's lifestyle such as reduction in sodium intake, exercise, and cessation of smoking are critical in the management of DKD[52]. Education and follow-up of patients is essential to guarantee compliance with treatment regimens in addition to complication tracking.

To summarize chronic kidney disease is one of the many complications that arise from unmanaged diabetes and it greatly contributes to morbidity and mortality of affected individuals. Regular comprehensive screening coupled with strict management of co-occurring pathologies can help identify the problem early on, influence the development of the condition, and improve overall results. The introduction of new pharmacological treatments, especially SGLT2 inhibitors and GLP-1 receptor agonists, have greatly changed the paradigm in the treatment of DKD and bring fresh optimism. With targeted treatment on both kidney and cardiovascular risk factors, the burden of DKD on health care and the quality of life of diabetic patients has the potential to dramatically improve[27].

## 2.6 Biomarkers of CVD and CKD

Managing illnesses like cardiovascular disease (CVD) and chronic kidney disease (CKD) along with diabetes requires constant vigilance, which makes the use of biomarkers crucial. Disease management becomes more vital as there is an increasing set of complications related to diabetes. In the case of CVD, inflammation and damage to the heart muscle may be indicative to trauma and inflammation. These are crucial for understanding the cardiovascular risks and can be evaluated using certain biomarkers of inflammation such as C - reactive protein (CRP) and high sensitivity troponin (hs-Tn). Lipid profiles such as LDL, HDL, triglycerides, and Cystatin are quite useful when evaluating atherosclerosis and heart failure. Heart attack and stroke can also be predicted using other biomarkers such as soluble CD40 ligand (sCD40L) and adiponectin. Monitoring these risk indicators is crucial for bringing about the timely diagnosis which aids in avoiding severe outcomes. The same can be said for CKD – early signs of kidney damage like albumin in urine (albuminuria) as well as reduced kidney function are monitored through serum creatinine and created estimated glomerular filtration rate (eGFR). Even though creatinine markers are widely accepted, some markers such as Cystatin C are more reliable. Further recognition of Chronic Kidney Disease also includes other markers such as Neck Tie 1 (KIM-1) and Neutrophil Gelatinase-Associated Lipocalin (NGAL) which help provide a warning around these injury indices. Markers these provide a window of intervention which can greatly improve the outcome.

## 2.7 Environmental factors

Environmental issues are potent contributors towards the onset of diabetes, cardiovascular disease (CVD), and chronic kidney disease (CKD) while also interacting with genes and lifestyle choices. In the case of diabetes, physical inactivity and obesity tend to be the most important manifestations. Sedentary behavior, which is defined as a lack of moderate to vigorous physical activity, can lead to weight gain, lower insulin sensitivity, and ultimately increases the chances of developing Type 2 diabetes. Moreover, there is evidence for the role of air pollution as an environmental risk factor, which correlates long-term exposure [22] to particulate matter and other pollutants with greater insulin resistance and an increased risk of Type 2 diabetes.

Concerning CVD, nutrition, inactivity, and smoking are critical risk elements. Diets containing high levels of saturated fats, Trans-fat, and refined sugar can elevate cholesterol and promote atherosclerosis which increases the severity of heart attacks and strokes. Heart disease is also caused by physical inactivity, which deteriorates the heart's strength and increases obesity. Moreover, smoking is recognized as one of the most powerful environmental risk factors related to CVD due to the damaging effects of tobacco on the blood vessels, increased blood pressure, and higher rate of clot formation[53]. The exposure to air pollution has also been associated with a higher likelihood of suffering from heart issues because of chronic inflammation of blood vessels and heart tissues caused by pollution[54].

With regards to CKD, chronic exposure to environmental poisons, poor quality of food, and not drinking enough water are additional factors that worsen existing kidney damage. Renal impairment is associated with the presence of environmental poisons like pesticides and heavy metals[55]. Insufficient water consumption, combined with excessive salt, can bring about high blood pressure which is a prominent risk factor for renal disease. Not to mention, people who are obese and do not exercise are almost guaranteed to develop and/or advance CKD because of their higher odds of getting hypertension and diabetes, both of which are harmful to the kidneys.

To sum up, smoking, lack of exercise, air pollution, poor diet, along with exposure to toxins have a direct impact on the development of diabetes, CVD, and CKD. Addressing these lifestyle factors through proper diet, exercise, and avoidance of harmful pollutants can significantly reduce the risk associated with these chronic diseases[56].

## 2.8 Machine Learning

As applied to artificial intelligence, machine learning is one of the most rapidly developing disciplines owing to its successful application in analyzing complex data sets across a massive range of fields such as healthcare and finance [57]. The central concept of machine learning involves enabling systems, with no instructions for a set task, to utilize the data provided towards making decisions, predicting outcomes, and improving their performance. This review focuses on the evolution, technical aspects, applications, as well as the challenges and prospective measures regarding machine learning.

### 2.8.1 Evolution of Machine learning

Machine learning can be traced back to the mid-20th century marked by the development of the branches of automated information processing by visionary like Alan Turing and John McCarthy. As far back as the 1950s and the 1960s, there was a clear focus on symbolic AI, whereby machines were furnished with certain rules deals. Of note, this method had several shortcomings, including managing uncertainty and complex data, which gave rise to the development of machine learning algorithms[58].

During the 1980s and 90s, the area underwent a paradigm shift with the application of statistical techniques in learning algorithms. Geoffrey Hinton, Yann LeCun, and Yoshua Bengio have been instrumental in working on neural networks, which resulted in the evolution of deep learning; a subset of machine learning concerned with artificial neural networks with many layers[59]. The field has seen rapid progress with the improved computing power and the availability of large datasets, with the deep learning models achieving best results in image recognition, natural language processing, and even playing games[60].

**2.8.2 Types and techniques of machine learning**

There are two major approaches to machine learning: supervised learning, unsupervised learning, and reinforcement learning.

a) **Supervised Learning**. In this approach, the model is trained on labeled data and learns to predict the outcome based on that label, thus the input is already mapped to a known output. Some of the algorithms in this category include decision trees, support vector machines (SVM), KNN (K nearest neighbors), and neural networks. In supervised learning, the recommendation systems or spam filters are standard examples; another application is regression (for example, predicting the price of a house)[61].

b) **Unsupervised Learning:** An unsupervised learning system uses the algorithms without providing any labels to it, so the system must try to find relevant patterns and structures using the given data. Clustering and association are the main techniques of unsupervised learning. Clustering techniques like k-means and DBSCAN combine data points that are similar, while association techniques like Apriori help in seeking the interrelations of data points in wider databases [6]. Unsupervised learning is used for customer segmentation and anomaly detection, among others.

c) **Reinforcement Learning:** Interaction with the surrounding is how one learns in RL. As in traditional learning, an agent performs different factions within an environment in RL and depending on the result (reward or penalty), the agent either progresses or regresses[62]. In most cases, the agent will eventually act in a manner that leads him or her to receiving the highest total reward possible and in turn utilize learning in carrying out the desired action. RL is well known for its successful applications in robotics, game playing (for instance AlphaGo), and self-driving vehicles.

### 2.8.3 Applications of Machine Learning

These days, multiple industries use Machine learning, and healthcare is no exception, as ML models come in handy for diagnosis, drug invention, and tailored medicine[60]. For example, some algorithms examine medical pictures for manifest tumors, while others analyze electronic health records in order to determine probable outcomes[63].

In finance, the application of ML includes fraud detection, algorithmic trading, and credit scoring. Fraudulent behaviors can be detected, and stock prices can be predicted using historical data with the aid of ML algorithms that evaluate tremendous amounts of transaction data[64].

In natural language processing (NLP), ML has transformed machine translation, sentiment analysis, and speech recognition. At the core of advanced systems like Google's BERT and OpenAI's GPT lie deep learning models such as transformers and recurrent neural networks (RNNs)[65].

To summarize, machine learning has undergone important developments in the past few decades, having its impact felt across different fields. With continuous research development and the introduction of new tools and practices, ML is likely to transform even more aspects of society and technology while providing solutions to pressing challenges. Still, ensuring a balanced approach toward ethical and interpretability challenges will determine its safe and responsible use in the future.

## 2.9 Role of machine learning in diagnosis of Diabetes mellitus

With the increased use of machine learning technology, there has been a significant improvement in accuracy, efficiency, and personalization in the diagnosis and management of diabetes and its comorbidities. Diabetes is a metabolic disorder, which means the body does not use blood sugar correctly, leading to hyperglycemia and severely increases the risk of complications, such as heart disease, nerve damage, vision loss, and kidney disease[45]. While older predictive methods remain efficient, their time-consuming nature and high probability of human error remains a flaw. Algorithms that have previously been implemented include supervised learning methods such as logistic regression, decision trees, random forests, and support vector machines (together deemed 'traditional ML'), all of which have been extensively researched and applied to predict and diagnose diabetes with electronic health records, demographic data, and biomarkers such as fasting blood sugar and Body Mass Index[38]. Importantly, other forms of deep learning, CNNs and RNNs, have greatly improved diagnostics in healthcare, and have simplified the evaluation of complex datasets, making them invaluable for analyzing the retinal images needed for diabetic retinopathy screening [14]. Diabetes outcomes have been shown to be more precise with the use of these methods and more advanced ensemble learning methods that utilize several models[66]. Not only for diabetic patients, ML has been critical in predicting comorbidities related to diabetes.

The integration of clinical data such as blood pressure, cholesterol, and smoking status enable the use of ML prediction models to accurately predict cardiovascular disease, a common comorbidity [67]. ML models have also been used to predict the onset of diabetic nephropathy by analyzing urine and blood biomarkers which allows for early stage preventative measures to be taken to slow disease progression. Deep learning algorithms have been trained to detect diabetic retinopathy through the analysis of retinal images, which has improved treatment opportunities and reduced the risk of vision loss. Nevertheless, obstacles remain regarding data quality and availability since ML models require large datasets, which remains the case in most healthcare environments. Furthermore, many ML algorithms, particularly deep learning models, are notorious for their opacity, which causes concern of trust and interpretability for clinicians. It also invites ethical and privacy challenges, especially around patient data, which requires stronger frameworks to ensure legal compliance with regulations such as GDPR and HIPAA. Aside from that, the incorporation of ML tools to clinical practice involves overcoming logistical, technical, and cultural barriers, including push back from clinicians.

Advancing research in this area should aim at designing clear and comprehensible machine learning models, accommodating multimodal data such as genomic data and data from wearables, and promoting collaboration between researchers, clinicians, and policymakers to resolve ethical and regulatory issues. Using machine learning capabilities, healthcare systems would be able to provide earlier diagnoses, tailor treatment to individual patient needs, and enable better health outcomes for patients suffering from diabetes, which would revolutionize diabetes management and its associated comorbidities.

## 2.10 Operation of Machine Learning

Machine learning (ML) works through a specific sequence of actions that allows certain algorithms to identify some patterns in data and make reasonable predictions or choices autonomously. The process usually starts with data collection where all relevant data sets are procured from, in this case, electronic health records, sensors, and survey forms. In the diagnosis of diabetes, for instance, data may include patients' socio-demographic details, laboratory test results, and other relevant activities. Data cleaning comes next, a stage in which data is prepared for analysis. This encompasses cleansing the data, strategically dealing with missing data points, normalizing or scaling features, and encoding categorical variables to prepare the data for ML classifiers.

When the data set is ready, it must be divided into testing and training components. The training set is the part of the data that the ML model learns from. It contains features (e.g., blood glucose levels) and a target variable (in this case, diabetes diagnosis). Some of the works-in-progress algorithms at this phase are supervised learning models, such as logistic regression, decision trees, and support vector machines, alongside the unsupervised learning model, such as K-mean clustering which assists in establishing some strange hidden patterns[68]. The model poses numerous equations which constituting its structural form or description the input variables, and during training, it progressively alters some of the parameters (weights) of these equations aimed at reducing the deviations of the actual output from the model's predictions of the output responses. This control is usually done using optimization methods such as gradient descent.

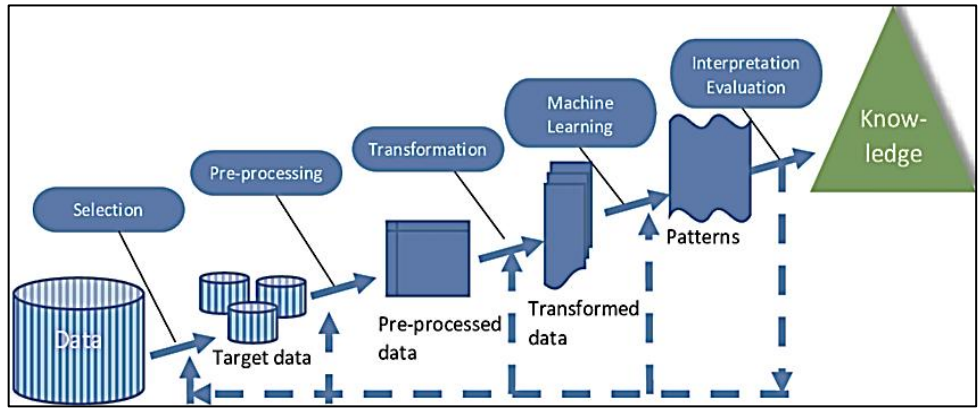

*Figure 2.2 Machine Learning Model*

*Source: Adapted from Johannes Hüffmeier et al. Trim and Ballast Optimization for a tanker based on machine learning - Scientific Figure on Research Gate. (Accessed 29 Apr 2025)*

To assess the performance of the model, it is evaluated with the testing set after training. The model's ability to generalize to unseen data is evaluated with metrics like accuracy, precision, recall, and F1-score. A model that performs well can be deployed in real-world applications, like predicting the risk of diabetes or diagnosing its comorbidities[69]. Nonetheless, ML models generally need a round of fine-tuning through hyper parameter optimization, along with cross- validation, to improve performance [70]. Also, models need to be maintained and adjusted over time to account for changes in data patterns, ensuring that they are accurate and reliable. The iterative cycle of training, evaluation, and deployment constitutes the primary activities in machine learning which enables it to tackle sophisticated challenges in numerous fields, including healthcare.

## 2.11 Supervised Learning

Supervised learning describes one of the most common and well-researched methods in which machines learn how to associate input data and desired output labels through algorithms trained on a labeled dataset. Its use is particularly common in prediction and classification tasks throughout different fields such as medicine, finance, image processing, and even language recognition[4]. In supervised learning, a model is built using a set of input variables (demographics, lab results) and an outcome (disease diagnosis), so that the model can be trained to relate the outcome to the provided input. In this article, I review the most important features, algorithms, applications, and issues of supervised learning based on recent literature [29, 38, 48, 71].

Learning algorithms can be divided into two categories: supervised and unsupervised. Classification problems attempt to predict one or more discrete labels whereas regression problems estimate the solution over a continuous domain. For instance, a patient diagnosed as having diabetes in one example would receive a label in binary classification (or yes/no) form. Classification tasks include identifying a patient as mild, moderate, or severe in a multi-class classification problem. Logistic regression, decision trees, random forests, support vector machines (SVMs), and neural networks are popular in classifying.

For instance, the classification of patients with diabetes using decision trees and random forests, and features such as age, BMI, and blood glucose were used to train the models[34]. In contrast with classification regression, algorithms work with continuous output. Some elementary examples include estimating blood sugar levels or projecting disease advancement throughout time. Popular regression algorithms include linear regression, ridge regression, and neural networks[45]

In recent years, focused advancements in supervised learning has been spent on the accuracy and interpretability of models. Ensemble techniques like gradient boosting (XGBoost and LightGBM) or bagging (random forest), which merge many weak learners to form one strong predictive model are becoming popular [45]. In addition, a multitude of specialists has transformed the domains of imaging and speech recognition through deep learning, one of the most examined branches of supervised learning, along with the application of the convolution and recursive neural network (CNN, RNN) architectures. Rrecent studies proved the remarkable accuracy of CNNs in recognizing diabetic retinopathy from retinal photographs with results that were similar to professional physicians.

Supervised learning has routinely been observed in the healthcare field for the more user-friendly and sophisticated diagnosis, prognosis, as well as a favorable treatment recommendation[21]. For instance, one study incorporated lifestyle and clinical features to predict cardiovascular disease in diabetic patients with supervised learning and was able to attain high precision [67]. In cancer treatment, supervised learning approaches have also been adopted to classify cancer types based on their genomes, which allows devising individual treatment strategies [72]. Outside of medicine, supervised learning techniques have been applied in credit scoring in finance, user segmentation in marketing, and linguistics for sentiment detection and machine translation [73].

The continued application of supervised learning techniques is met with various obstacles. One of the unfortunate traits of supervised learning is the reliance on labeled data, which is often costly and can take a long time to obtain. This is particularly true in such fields as medicine, where expert level annotation is required[1]. Additionally, supervision models are particularly vulnerable to parameters estimation, since they do well for training data but fail with unseen data. There have been algorithms developed, such as cross validation and regularization, and dropout, commonly used with neural networks, which attempt to solve the problem[28]. A further challenge is interpretability of the model, particularly with more advanced models like deep neural networks which, unfortunately, function as 'black boxes'. Work in this area includes XAI techniques such as SHAP and LIME which attempts to explain the model's decisions[74].

**2.11.1 Logistic Regression (LR)**

In Logistic Regression, a statistical approach for predicting a target variable's value is 0 (No Disease) or 1 (Disease Present). A logistic function is applied, and the output falls within a score ranging between the boundaries of 0 and 1. If the sigmoid probability is higher than the defined threshold (commonly set at 0.5), it will be classified as "disease present," and thus LR model will predict the disease. Owing it its simplistic nature, history of linear disease prediction models based on LR is very rich[48]. Its clear structure and high computational efficiency makes it a strong candidate for structured medical datasets. However, it has limited flexibility when it comes to covering intricate patterns due to the relationships between independent variables and the log-odds of the outcome being strictly linear [37]. Among other cases, it has been successfully utilized in prediction risk of CVD and CKB, primarily when clinical risk factors of blood pressure and cholesterol levels, along with lifestyle habits, are integrated into the model [38].

### 2.11.2 Random Forest (RF)

An algorithm known as Random Forest (RF) relies on a set of trees, or an ensemble of learning trees, where each tree is a decision tree, and each one delivers a classifier. To improve accuracy and reduce overfitting, RF constructs multiple decision trees and aggregates their predictions to improve accuracy and reduce overfitting[75]. Each decision tree is trained on a random subset of the dataset (bootstrapping) with a different random selection of features, ensuring diversity among the trees[76]. RF gives predictions via majority voting in case of classification tasks and averaging in regression contributions. RF provides robust resistance against noisy and imbalanced datasets which is a common problem in healthcare data that contains missing values or class imbalances[47]. In addition, RF features scores which support identifying key risk factors for RF. Although RF does provide ample benefits, it is computationally expensive when tasked with large datasets, and its complexity makes it less interpretable than simpler models like LR, which is linear regression. Studies have shown that RF performs adequately in CKD classification and CVD prediction, especially when clinical, demographic and lifestyle features are combined with other features[77].

### 2.11.3 Support Vector Machines (SVM)

Support Vector Machines (SVM) is one of the supervised learning models that gives good results in classification problems by determining the optimal hyperplane for separating the classes in a given classification problem[78]. When the data is not linearly separable, SVM uses the kernel trick to map the data into a higher dimensional space where a linear separator can be used. Several types of kernels which can be used include linear, polynomial, and radial basis function (RBF) kernels[47]. SVM is particularly beneficial when analyzing high dimensional datasets. This is especially true in the domain of medical imaging, where the complexity of data is substantially high. It also performs well in disease classification tasks such as CVD and CKD when an amalgamation of numerical and categorical data is present [79]. On the downside, SVMs tend to take a longer time to train especially when there are large amounts of data, and they are sensitive to hyper parameter tuning (for example, the kernel versus regularization parameter ratio). Numerous prior works have shown that SVM performs best in detecting CVD from ECG signals and CKD from clinical biomarkers[80].

## 2.12 Gap Analysis

There are still significant gaps that inhibit machine learning's potential in South Asia, especially within clinical practice. One notable hindrance is the scarcity and heterogeneity of datasets, particularly within chronic kidney disease (CKD). Public datasets readily available within Pakistan suffer from a disproportionate focus on diabetes and cardiovascular diseases (CVD), largely ignoring CKD cases[27]. The Buner CKD dataset is one of the few available resources within the country; however, it is not freely accessible which limits reproducibility, external validation, and prospective longitudinal studies[76]. In addition, numerous regional datasets such as the Heart Failure Clinical dataset (299 samples) and Buner CKD (500 samples) are dwarfed by larger datasets from other regions, which diminishes model generalizability and increases the likelihood of overfitting. This highlights the necessity for diverse datasets that can be openly accessed to build reliable predictive models[79].

Furthermore, the lack of integrated or comparative frameworks of South Asian countries presents a significant gap. Given the shared demographic characteristics and disease burden within the region, the majority of studies tend to focus on single-country evaluations, which hampers the formulation of generalized approaches that can be more widely applicable[35]. Another important overlooked area is the insufficient interpretability of models. Numerous studies boast accuracy metrics but do not empirically justify the model's reasoning using relevant medical features with appropriate statistical reasoning[59]. This failure to explain restricts clinical acceptance and incorporation. Resolving this issue by using machine learning combined with rigorously defined statistical feature validation is necessary in order to enable the development of reliable, accountable, and clinically relevant AI applications for healthcare in South Asia[80].

# Chapter 3
# METHODOLOGY

## 3.1 Study Design

This study used cross-sectional analytics research design to assess the possible early identification of chronic kidney disease (CKD) and cardiovascular disease (CVD) in diabetes patients using conventional statistical methods as well as modern machine learning techniques. The objective of the study was to evaluate whether physical and biochemical features included as clinical markers differentiated patients grouped by disease stage, and whether selected features utilized for prediction modeling were sufficiently distinctive. The study was carried out within five months from May 2024 to September 2024 at one of the tertiary care hospitals in Karachi, Pakistan, using stratified random sampling.

## 3.2 Sample Size

To determine the sample size open EPI website calculator was used for estimating population proportion. The formula is

$$n = \frac{Z^2 P(1-P)}{d^2}$$

Where,

Z= confidence interval at 95% (standard value of 1.96)

P= prevalence of disease in project area

d = margin of error at 5% (standard value of 0.05)

*Table 3.1 Sample Size Calculation Using Open Epi Formula*

| Parameter | Value | Description |
|---|---|---|
| **Confidence Level (Z)** | 1.96 | Z-value for 95% confidence interval |
| **Prevalence (P)** | 0.267 | Estimated diabetes prevalence in Pakistan (26.7%) |
| **1 - P (Q)** | 0.733 | Proportion of population without diabetes |
| **Margin of Error (d)** | 0.05 | Desired precision (5%) |
| **Calculated Sample Size (n)** | 301 | Minimum number of participants required |
| **Actual Sample Size Used** | 703 | Final number of diabetic patients enrolled in the study |

As per the IDF Diabetes Atlas 2021, diabetes prevalence (P) in the adult population of Pakistan is estimated at roughly 26.7%. This is based on a 95% confidence level and 5% precision interval. For optimal sample size estimation as shown in Table 3.1, at least 301 participants would have sufficed, but this study aimed for greater reliability and included 703 diabetic patients allowing for CKD only, CVD only, both, and subgroup analyses alongside enhanced statistical potency.

## 3.3 Selection of subjects

From the Outpatient Department (OPD) of the hospital, a total of 703 patients were recruited. Both male and female patients with a prior diagnosis of diabetes within the ages of 18 and 75 were included as participants. Data collection involved the use of a structured questionnaire which contained a section for each patient's profile, a section for their medical history, a section on the clinical signs and symptoms of the patient, as well as a section on laboratory tests conducted. Patient data and their clinical reports were double-checked against the patient's self-reported answers to the possible extent. Under the supervision of a medical professional. Anonymized data was securely stored for each participant.

**3.3.1 Inclusion Criteria**

Participants were chosen for this study as per the following stipulations:
- Diabetes mellitus patients diagnosed with type II and aged above 18 years.
- Complete clinical and biochemical data including, but not limited to, HbA1c, serum creatinine, and lipid profile must be available.
- Patients attending the Outpatient Department (OPD) within the timeframe of this study (May–September 2024).
- Participation in the study with the ability to withdraw consent at any time before or during collection of identifiable data.
- Classified into one of the following disease categories:
  No CKD/CVD, CKD only, CVD only and Both CKD/CVD

**3.3.2 Exclusion Criteria**

The study did not include participants who met the following criteria:
- Pregnant women.
- Participants lacking comprehensive laboratory or clinical information.

**3.3.3 Method of data collection**

The data gathering process for this cross-sectional study was done between May to September of 2024 at the OPD of one tertiary care hospital in Karachi city Pakistan. The study's participants were patients suffering from diabetes mellitus with their previous clinical records accessible Furthermore, the clinical experts for chronic kidney disease (CKD) and cardiovascular disease (CVD) helped construct the parameters necessary for the chronic kidney disease (CKD) and cardiovascular disease (CVD) classification, diagnosis, and risk assessment to be securely encompassed.

The developed questionnaire was divided into three major subheadings:

(1) Completion of a socio-demographic questionnaire;

(2) Past medical and laboratory test history;

(3) Informed consent.

**3.3.4 Performa**

To methodically obtain all relevant information from participants, a custom health survey performa was produced. The performa was sectioned out into personal information, medical history, and consent. *(Full performa is placed at Appendix A of the thesis. It contains all questions and fields used for this study.)*

**3.3.5 Ethical Considerations**

The ethical clearance and oversight for this study was reviewed by the Creek General Hospital Institutional Review Board (IRB) and was given approval. All the data utilized were anonymized, and where required, basic informed consent was obtained, thereby ensuring confidentiality as mandated by the applicable laws and regulations.

**3.3.6 Grouping of Subjects**

Based on the clinical diagnoses, participants were organized into four groups:
- Group A: Diabetes patients with both CKD and CVD (n = 82)
- Group B: Diabetes patients only with CKD (n = 220)
- Group C: Diabetes patients only with CVD (n = 75)
- Group D: Diabetes patients without CKD or CVD (n = 326)

This approach allowed the study to conduct group specific statistical analyses and create customized machine learning models for each disease state.

**3.3.7 Biophysical and Biochemical Parameters**

The following biophysical parameters were assessed for every patient in the study:

1. Age (Years)
2. Gender (Male/Female)

CHAPTER NO. 3                                                                                                   METHODOLOGY3. Duration of diabetes (years)
4. Body Mass Index (kg/m²)
5. Blood Pressure: Systolic/Diastolic (mm/Hg)
6. History of Myocardial Infarction and Stroke (yes/no)
7. Known case of Hypertension (yes/no)
8. History of Kidney and Heart Disease (yes/no)

The following biochemistry and laboratory parameters are obtained from the medical records or the most recent laboratory report:

1. HbA1c (%or mmol/mol): This biomarker indicates the long term glycemic control.
2. Serum Creatinine (mg/dL): This marker indicates renal function.
3. Serum Urea (mg/dL): This indicates secondary renal function.
4. Total Urinary Protein (mg/dL): Presence of protein in the urine either qualitative or quantitative result.
5. Cholesterol (mg/dL)
6. Triglycerides (mg/dL)
7. Troponin: (ng/mL) this is measured if available as an indicator of cardiac injury.

### 3.3.8 Determination of Body Mass Index

Each participant's height and weight were taken with a calibrated stadiometer and digital scale. Weight was captured in kilograms (kg) and height in centimeters (cm). For calculating BMI, height was translated from centimeters into meters (m).

BMI was computed applying the Quetelet's Index formula:

$$BMI = \frac{Weight(kg)}{Height(m)^2}$$

Subsequently, each participant's BMI was stratified by WHO standards which can be seen in Table 3.2.

*Table 3.2 Body Mass Index (BMI) Classification*

| BMI Range (kg/m²) | Classification |
|---|---|
| < 18.5 | Underweight |
| 18.5 – 24.9 | Normal weight |
| 25.0 – 29.9 | Overweight |
| ≥ 30.0 | Obese |

This classification supported the investigation of body composition and the impact of CKD and CVD in diabetic patients. Increased BMI is widely accepted as a contributing factor for both renal and cardiovascular complications and served as a principal factor in this investigation. This research focused on two distinct phases:

**(a) Phase I: Statistical Profiling and Feature Selection**

This phase consisted of an independent literature review, medical expert consultation, and statistical analysis in SPSS (v27.0). A diabetic population with CKD and CVD comorbidity was confirmed through identifying key features during the literature review. Machine learning modeling was done using significant parameters ($p < 0.05$) identified through ANOVA tests that included serum creatinine, HbA1c, cholesterol, stroke, myocardial infarction, BMI, and hypertension as inputs.

**(b) Phase II: Machine Learning Modeling**

During this phase, the data obtained from the questionnaire and medical files was applied to create and assess machine learning algorithms for predicting CKD and CVD in diabetic patients. The models used were Logistic Regression, Support Vector Machine, and Random Forest. To balance classes, SMOTE was implemented, and standard model evaluation metrics such as AUC, accuracy, precision, recall, and F1-score were computed alongside other benchmark evaluations.

## 3.4 Statistical Analysis

Data were processed with IBM SPSS Statistics (Version 27) applying these methods, Continuous variables (mean, standard deviation, standard error) and categorical variables (frequencies, percentages) were summarized using descriptive statistics.

One-way ANOVA was used to assess the differences in means across four disease groups (No CKD/CVD, CKD only, CVD only, Both).

- Levene's Test was used to check for equality of variances.

- Significant pairwise differences were tested using post hoc tests Tukey HSD and Bonferroni.

- The eta squared ($\eta^2$) values for effect size estimation were calculated for every variable.

- P-values ≤ 0.05 were considered statistically significant.

## 3.5 Machine Learning Algorithm Design

This sub-section outlines the methods used in the study along with prior steps of data preparation, implementations, and models evaluation. With respect to the early identification of cardiovascular (heart) and kidney complications in diabetic patients, three supervised machine learning algorithms are used: Logistic Regression (LR), Support Vector Machine (SVM) (RBF kernel) and Random Forest (RF). Since diabetes poses one of the greatest risks for chronic kidney disease (CKD) and cardiovascular disease (CVD), detecting these complications early on is important for precise action. Every aspect of the methodology, including data processing, validation, and the development of evaluation benchmarks, is tailored for this objective. Every experiment was done exclusively in Python 3 using recognized machine learning packages like Scikit-learn, Imbalanced-Learn, and Pandas; the experiments were conducted according to IEEE guidelines for reproducible research.

In the following sections, the setup of the environment will be described, then data will be described, including class imbalance mitigation, model training with stratified cross-validation and evaluation techniques for the main implementation.

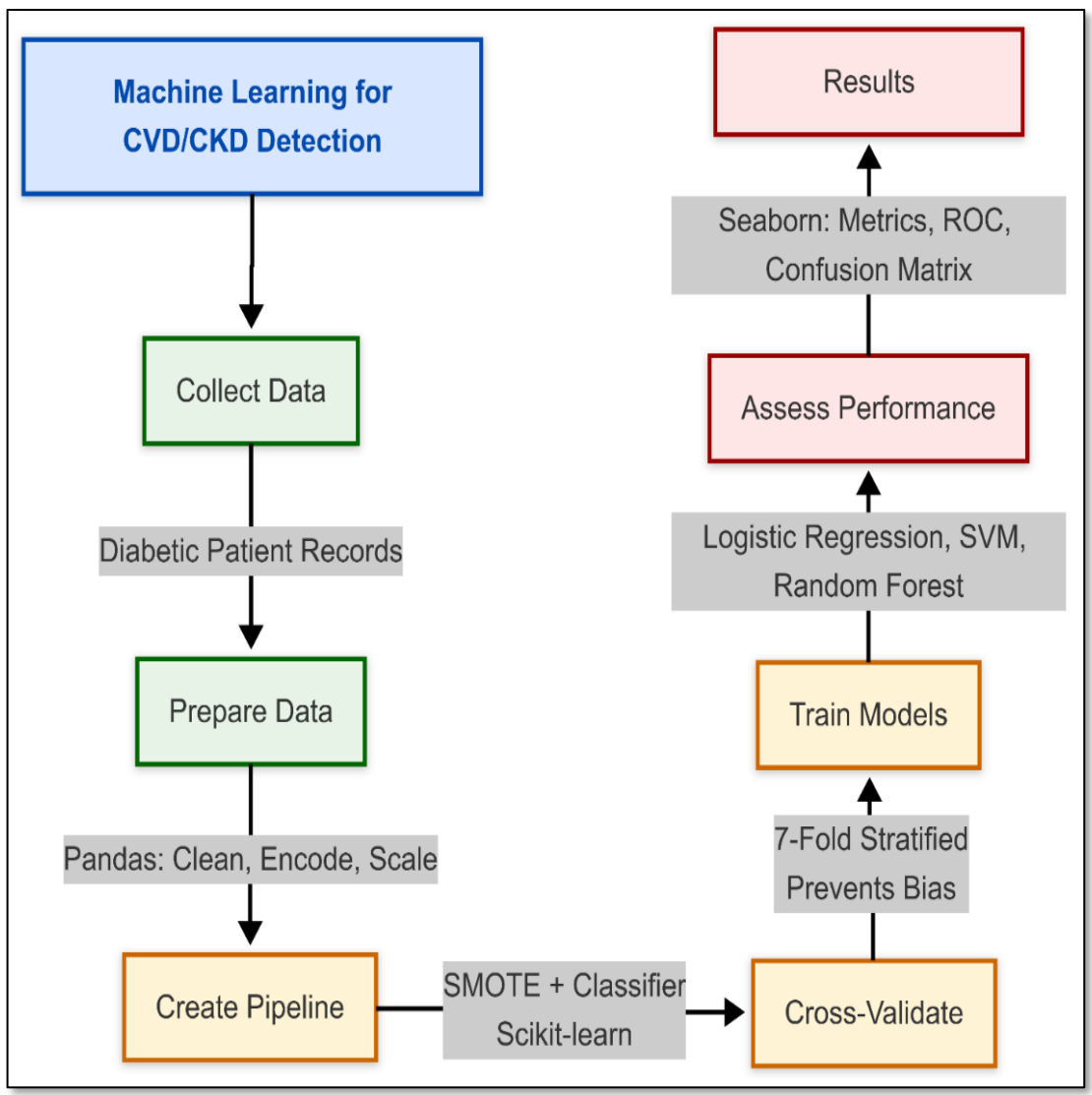

*Figure 3.1 ML model for predicting CKD and CVD*

**3.5.1 Experimental Environment and Tools**

Running on Jupyter Notebook, Python 3.8 formed the development environment. Included key libraries and their versions: Pandas for data manipulation, NumPy for numerical calculations, Scikit-learn for model training and evaluation, Imbalanced-learn for oversampling techniques (SMOTE), and Matplotlib/Seaborn for visualization. Above mentioned library imports are used all across the pipeline and installation of necessary dependencies. Pandas read Excel files using the openpyxl engine. Relevant functions were set with a fixed random seed—e.g., random_state=42 to guarantee result repeatability.

**3.5.2 Dataset Loading and Preprocessing**

This study's dataset comprises the medical records of diabetic patients that contain features associated with their heart and kidney health, such as blood glucose level, blood pressure, cholesterol, indicators of kidney function, etc., and an outcome label signifying if the patient has early signs of heart and/or kidney disease. The data was downloaded from a Microsoft Excel file (.xlsx) format.

The Excel file was imported into a Pandas Data Frame. This displays the shape (number of rows and columns) of the dataset as well as the first few records to check that the dataset was loaded correctly. The dataset consists of a number of predictor features (columns) and one or more target columns. For the current research task, this study uses a binary classification target where the patient is ailing with (or at early onset risk) heart or kidney disease. If there were multiple outcome columns available (for example, one for heart disease and one for kidney disease), individual models could be constructed for each; however, this research opted for a single outcome in our pipeline and therefore used a binary label indicating whether the patient suffers from either complication (in our case, ComplicationPresent: yes/no).

Data Cleaning: As a part of quality assurance, basic preprocessing was performed which covers missing value treatment (like eliminating records with excessive missing fields or imputing values where suitable) as well as the removal of outliers that appears to be inaccurate. Consider for example the case where any feature has impossible values such as negative values for age. These would have to be addressed. In the current dataset, the amount of cleaning needed is minimal because the data had already been organized.

Feature Selection: All relevant clinical and demographic features were included in the model. Removal of non-informative identifiers like patient ID was done, which is simply a number and does not aid in prediction.

Encoding Categorical Variables: The dataset had some categorical attributes like Gender, Known Case of diseases and other similar types. All features were transformed to a numerical value due to requirements of machine learning algorithms. Hence the variable was encoded. Nominal categories will be one-hot encoded while ordinal categories are encoded ordinally. For instance, to do one-hot encoding, we employed Pandas' get_dummies or Scikit-learn's LabelEncoder/OneHotEncoder as appropriate. This research detected object-type columns as categorical and apply pd.get_dummies, which splits each categorical column into several binary columns (one for each category level with drop_first = True). This process ensures the information is in numerical form. Consider the column Gender with male and female as values, it will be replaced by a binary column Gender, where Female is implied by a value of 0 and male by 1.

Feature Scaling: Continuous features underwent scaling before modeling in order to adjust their intervals. Scaling is to address some factors more critical than others (SVM and logistic regression) in which features with a bigger numeric value do not dominate the distance computations. Scaling was done using standardization (zero mean, unit variance) via Scikit-learn's StandardScaler. Scaling was done after encoding the categorical variables to ensure all features are numeric.

ComplicationPresent is likely the assumed binary target column (0 = no complication, 1 = complication present). This feature X was extracted and the remaining feature matrix was scaled. The scaler's fit_transform is applied to the full feature set for ease of explanation; however, keep in mind that during the cross-validation phase, scaling, alongside other preprocessing steps such as encoding and SMOTE, is restricted to the training folds for encasing information leakage from the test folds.

### 3.5.3 Handling Class Imbalance with SMOTE

In real-world medical datasets class imbalance is a frequent occurrence. The medical condition of interest (in this context, early stage heart/kidney disease) is quite rare as compared to the negative class 'no complication.' In this research's dataset, diabetic patients had early signs of CVD/CKD only in a much smaller fraction compared to those who did not, which created an imbalanced class distribution. A dataset is termed as imbalanced if the classification categories are not roughly equally represented. In this research, a classifier that is trained on this data will achieve high accuracy (biased toward the majority class like 95% of patients diagnosed "no") simply by predicting majority class: i.e. "no complication" while failing to detect the minority (positive) cases: "with complication" (it considers true cases as false, which is unacceptable for early medical diagnosis i.e. the false negative is very costly). Moreover, Figure 3.2 gives a clear representation on the flow of preprocessing dataset.

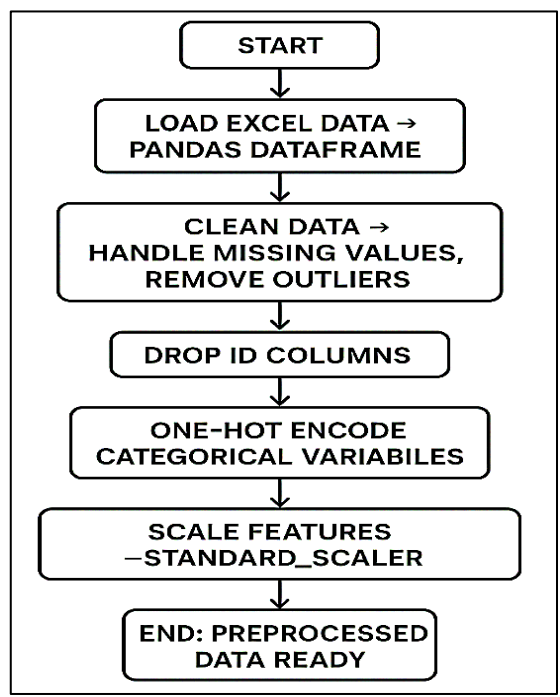

*Figure 3.2 Preprocessing of the dataset*

Moreover, in this study the classes were balanced in training data by applying SMOTE (Synthetic Minority Over-Sampling Technique). This method is effective in oversampling a specific class like the "Positive" class. It does this by creating synthetic instances of the minority class instead of just creating copies of the minority instances. Providing more data for the positive class enables the classifier to learn the decision boundary effectively for the "Minority" class.

Applying SMOTE exclusively to the training section of the data is necessary during model validation to prevent any potential data leakage. If oversampling is performed on the entire dataset prior to cross-validation, it will lead to optimistic model performance because information from the test-fold would influence training through artificially generated data. Hence, in the pipeline, SMOTE was incorporated with cross-validation to ensure during each fold, the training portion of the split is oversampled while the validation portion is left undisturbed to mimic real-world scenarios with class imbalance.

In practice, this was accomplished by implementing SMOTE from the module 'Imbalanced-Learn' together with a pipeline from Scikit-learn, or within the cross-validation for-loop. One method was to create a Pipeline that first applies SMOTE, then fits the classifier and use this pipeline in cross-validation. Another method was to manually apply SMOTE on each fold in a for-loop. For simplicity, this study chose the pipeline option.

This research constructed a two-step pipeline: step one generates synthetic samples of the minority class using SMOTE; step two fits a logistic regression classifier. The cross_val_score function takes care of splitting the data into stratified folds and for each fold, fits the pipeline meaning that SMOTE will only be done on the training data of the fold for that fold internally). For illustrative purposes here, F1-score was selected as the metric because it is a better measure when precision, recall, and accuracy are vital components, especially in the case of evaluating imbalanced datasets. This research undertook a similar approach with the other models (SVM and RF), either creating distinct pipelines or looping through to apply SMOTE in a loop to each model.

This method of applying SMOTE alters the sensitivity of the classifiers to the minority class, which is important because in the healthcare domain, mistaking an early case of the disease for a healthy patient (false negative) is much worse than the opposite scenario. Adjusting cross-validation keeps the evaluation processes balanced regardless of the disparity, it still makes use of fair assessment techniques that maintain equilibrium despite the lack of proportionality.

**3.5.4 Stratified K-Fold Cross-Validation**

To evaluate the models, this research study used a Stratified K-Fold Cross-Validation technique. Cross-validation is the process of dividing the dataset into k parts (folds), followed by training the model k times (one for each part), where each part is designated as a validation set while the remaining parts are used for training. Stratified K-Fold attempts to maintain the class proportions for each subgroup used in training. This is particularly useful for class imbalance problems. In classification problems with imbalance classes, stratified sampling attempts to create representative samples by including all classes to provide a more accurate estimate of performance. In case of a dataset, if 20% of a patient population suffers from complications, stratification makes sure all folds have approximately 20% positives preventing a situation where a subset has no positive cases and thus making the accuracy misleading.

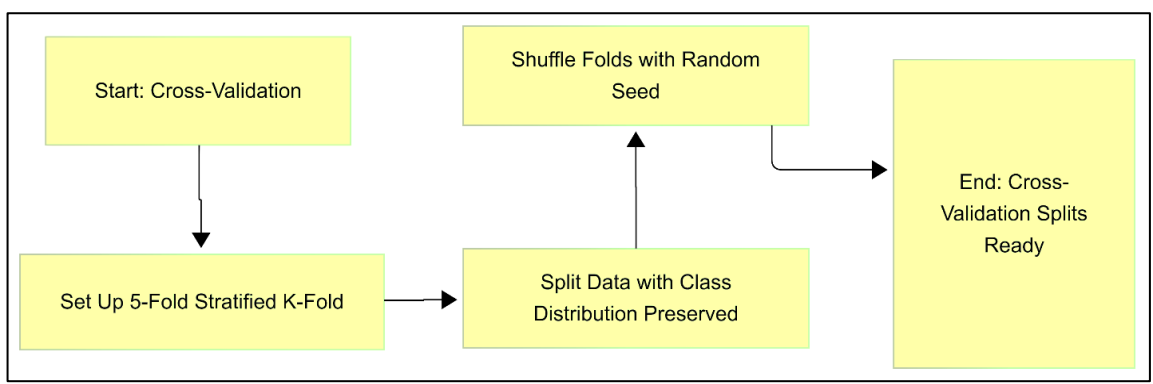

*Figure 3.3  Cross Validation*

To maintain a balance between computing efficiency and rigorous evaluation, this research settled on the methodology utilizing 7 fold stratified CV (i.e. k=7). Seven fold CV implies the dataset is split into 7 equal parts, where each model is trained 7 times, each time reserving one part for validation and training on the other six. When instantiating the StratifiedKFold object, this study set shuffle=True and used a fixed random seed to ensure reproducibility to some extent while also allowing randomness in fold composition.

Moreover using loop, this study obtained the training/validation sample counts per each fold and their respective positive case counts to validate stratification (each fold's positive count was approximately 2 times 20% of that fold). As described earlier, we implemented this CV splitting approach into the SMOTE with model pipeline but also utilized cross_val_score and cross_val_predict which are higher level abstractions of the loop in the background, but do similar processes under-the-hood. Figure 3.3 shows the flowchart of the cross validation.

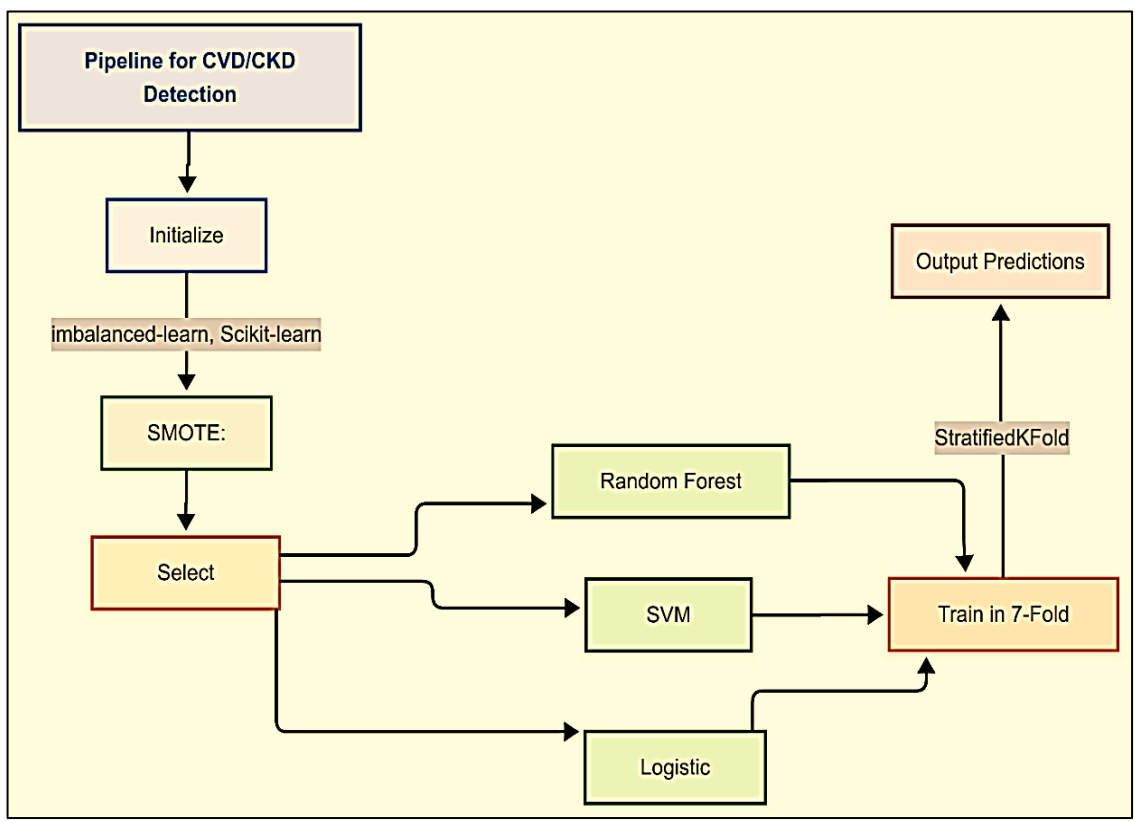

*Figure 3.4 Model Pipeline*

The advantages of using stratified K-Fold CV include:

(1) It reduces the variance associated with a given train-test division by performing cross-validation and averaging the results.

(2) It optimally utilizes infrequent datasets (each data example is served k-1 times in training and tested once)

(3) It maintains dependable evaluation metrics even in the presence of class imbalance due to stratification in samples.

This method corresponds to the best-known approaches for medical AI studies owing to the fact that data is often not uniformly distributed.

### 3.5.5 Machine Learning Models Implementation

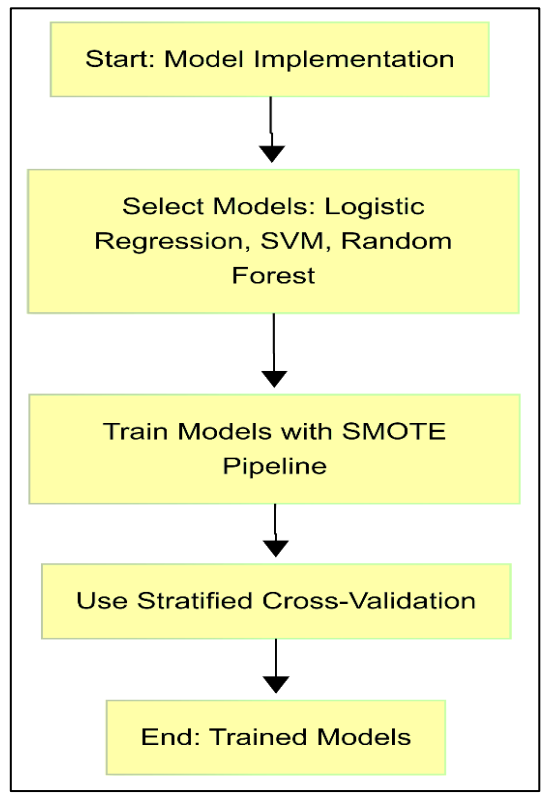

*Figure 3.5 ML Model implementation process*

This research study looked into three classification models: *Logistic Regression, Support Vector Machine (RBF kernel), and Random Forest.* The rationale behind these models was to include a linearly simple, powerful nonlinear, and an ensemble learning approach. Prior research work in prognosis has shown that logistic regression is often surprisingly effective and competitive with much more complex models, SVMs with proper kernel settings in high-dimensional data, and ensemble methods such as Random Forest perform well due to the treatment of interactions among features. Furthermore, this research study have outlined the execution steps of each model, as well as how they were trained with the previously outlined cross-validation scheme.

*Logistic Regression* is perhaps the most popular linear model for binary classification. Like other linear models, it attempts to model the log-odds of the positive class as some linear combination of the features. While it's a simplistic model with the decision boundary containing a straight line, it has proven to be a strong baseline in terms of performance. Logistic regression sometimes approximates the accuracy of much more complicated algorithms. The ability to interpret the outcomes through coefficients makes it defendable in front-line medicine where decisions have to be made involving critical risks.

This research study applied Scikit-learn's Logistic Regression with default settings, except maximum iterations was adjusted to ensure convergence. The C parameter which sets the inverse of the strength of regularization was not modified and left at the default 1.0 (L2 regularization) and no other tuning was conducted in this phase of the methodology. No explicit RFE (recursive feature elimination) was performed as our aim was to assess model performance using the available features; however, the outputs were scrutinized to determine the significance of the features.

The logistic model was trained within the stratified 7-fold CV. With the pipleline approach with SMOTE, each training within a fold first oversamples the minority class and then applies logistic regression on the respective fold's training data. This obtains 7 sets of model parameters (one per fold) and corresponding validation predictions. It is possible to cross validate and produce an estimate of generalization performance without requiring a separate hold-out set. Here, cross_val_predict trains the lr_model 7 times (as defined by skf) and returns the combined predictions for every participants in the validation fold. These can directly compared y_pred_cv_lr against the true labels y to compute metrics.

***Support Vector Machine (SVM)*** is a powerful classifier that finds an optimal hyperplane which maximally separates classes in a high-dimensional space. By employing kernel functions, SVMs efficiently perform nonlinear classification by mapping inputs into higher dimensional feature spaces.

This research study opted for the Radial Basis Function (RBF) kernel because it is widely used in nonlinear problems and appropriately triages situations where the linkage between features and the class label is complicated, as it is not linearly separable.

Scikit-learn's SVC (Support Vector Classifier) with kernel='rbf' was used. The other parameters were kept at defaults (C=1.0, gamma='scale'). These default RBF settings often constitute a sensible baseline for SVM. They would need to be adjusted carefully in more refined studies with grid search or random search for C and gamma values. For this methodology, the primary focus is on showcasing the pipeline, not performing precise adjustments.

Just like with logistic regression, the SVM was trained within a stratified CV loop. Unlike logistic regression, SVM training can sometimes take longer, especially with larger datasets which necessitated scaling features (which we did) to improve convergence speed and accuracy. Additionally, this research set probability estimates as True in SVC to obtain probabilistic outputs for ROC curve plotting which uses Platt's scaling internally after the model is trained.

For patients included in y_proba_cv_svm, this research study retrieved probabilities for early disease classified as class 1. This information was used for plotting the ROC curve that will be detailed in Chapter 4. Performance for SVM regarding classification metrics is similarly logistically evaluated by predicting cross-validated values y_pred_cv_svm = cross_val_predict (svm_model, X_scaled, y, cv=skf) and comparing against the true labels.

*Random Forest (RF)* is an ensemble learning approach that creates numerous decision trees and combines their outputs (using majority voting for categorization) improving overall performance. RF is famed for accommodating non-linear feature relationships and interactions. Additionally, RF is known to be robust overfitting due to the averaging of multiple trees. This research opted to include Random Forest because it has proven useful in a variety of classification problems in biomedicine and because it may be able to capture intricate patterns within the data that would be missed by a single tree or a linear model.

This research applied Scikit-learn's RandomForestClassifier with 100 trees (n_estimators=100) leaving the default values of the splitting criteria Gini impurity. Once more, no elaborate hyperparameter tuning was performed; the defaults, such as maximum depth, were retained which generally work well. The random forest handles binary and numeric features automatically, even without scaling, (being tree-based and scale-invariant) but it's still provided it with the X_scaled data.

The Random Forest was trained and evaluated with the 7-fold CV. As RF manages class imbalance issues through the modification of class weights, one could argue that the lesser represented class could be given a higher weight; for these purposes, this research relied on SMOTE for class balancing.

This research than computed precision, recall, F1-score for each class and total accuracy for the random forest performance, balanced for all folds discussed in chapter 4. The importance of features for the RF model can also be assessed by fitting on the entire dataset and checking rf_model.feature_importances_ , but that is beyond the methodology focus in this case, which is centered in the evaluation process.

**3.5.6 Evaluation Metrics and Model Assessment**

The research study applied a thorough assortment of evaluation tools accuracy, precision, recall (sensitivity), F1-score, and the ROC curve with AUC, to evaluate the predictive performance of the models in the context of early disease detection. In a medical prediction setting, every metric offers insight into several aspects of performance. For a more understandable evaluation, the study also used visual analysis tools including ROC curves and confusion matrices. To guarantee an objective assessment, all measures were derived from the combined cross-validation forecasts.

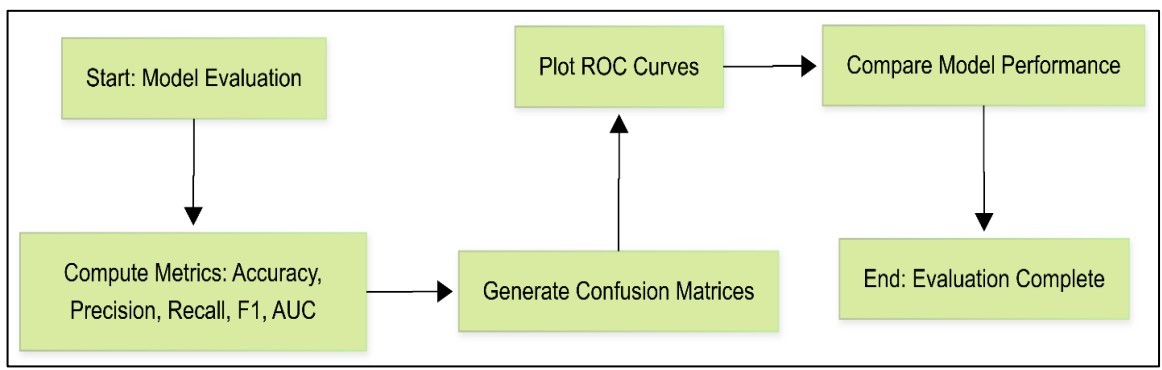

*Figure 3.6 ML model evaluation process*

Precision is the proportion of true positives (TP) to all projected positives (TP + FP). Recall (sensitivity) is the proportion of true positives to all actual positives (TP + FN). Recall is very crucial in medical screening; this study wants to catch as many genuine cases as possible (minimize false negatives). To prevent false alerts (false positives), precision is also rather crucial; yet, often in this field the price of a false negative is greater.

The harmonic mean of recall and accuracy is the F1-score. It offers a single indicator of a test's accuracy that balances both issues. An F1 nearer to one shows both great accuracy and great recall. This research study published class-wise accuracy, recall, F1 (for positive and negative classes) as well as the macro-averaged values (via classification_report from Scikit-learn).To see the number of true negatives, false negatives, false positives, and true positives for each model, this research study calculated confusion matrices. The confusion matrix offers a clear summary of how the predictions of the model are spread out. From the confusion matrix, for instance, the study directly showed how many false alerts were issued (false positives) and how many early disease cases were missed (false negatives) by showing counts in a 2x2 matrix. This study did this for every model to evaluate its strengths and shortcomings; for example, one model might have less false negatives but more false positives than another, suggesting a trade-off between recall and precision.

The Receiver Operating Characteristic (ROC) curve is a graph of the True Positive Rate (TPR or recall) versus the False Positive Rate (FPR) at several classification thresholds. It shows the trade-off between sensitivity and specificity. While a perfect classifier would hit the top-left corner (AUC = 1.0), a classifier that makes random guesses would generate a ROC curve along the diagonal line (AUC 0.5). A scalar summary of the ROC, the Area Under the ROC Curve (AUC) indicates the likelihood that the model ranks a random positive instance higher than a random negative one. Its threshold independence and relevance for imbalanced data assessment led this study to select ROC/AUC as a main measure. Using the cross-validated probability estimates for the positive class, the ROC curves were produced for each model.

Using the above measures, the study assessed each model's performance by averaging across the cross-validation folds. The metrics calculated (precision, recall, etc., from the confusion matrix or classification report) practically represent the micro-averaged metrics across all folds since the study applied cross_val_predict on the whole dataset. That is, they consider the gathering of all fold predictions and truths as one large evaluation set, which is a reasonable method to combine cross-validation outcomes. On the other hand, this research study also considered the per-fold metrics and then averaged them (using cross_val_score as indicated for F1 previously). In this research studies instance, both strategies produced comparable results.

# Chapter 4

# RESULTS AND DISCUSSION

## 4.1 Descriptive Statistics by Group

The descriptive statistics for all different variables of interest for the four patient groups in the study are summarized in Table 4.1. These groups include Group A Both CKD & CVD, Group B CKD only (n = 220), Group C CVD only (n = 75), and Group D No disease (no CKD or CVD, n = 326). For health indicators that are continuous in nature, means along with standard deviations (SD) were calculated; as for binary variables (gender and comorbidities), outcome proportions were calculated and presented in percentages.

*Table 4.1 Descriptive statistics for key variables by CKD/CVD group.*

| Variable | No Disease (*n=326*) | CKD Only (*n=220*) | CVD Only (*n=75*) | Both CKD & CVD (*n=82*) |
|---|---|---|---|---|
| **Gender (% Male)** | 31.0% | 68.2% | 34.7% | 75.6% |
| **Age (years)** | 52.21 (10.29) | 52.64 (14.55) | 50.59 (14.79) | 55.59 (13.95) |
| **BMI (kg/m²)** | 27.41 (4.42) | 25.46 (2.61) | 24.95 (5.33) | 24.99 (3.89) |
| **HbA1c (%)** | 8.62 (2.30) | 7.80 (2.39) | 9.22 (1.79) | 9.14 (1.92) |
| **Creatinine (mg/dL)** | 0.793 (0.483) | 2.274 (0.807) | 0.955 (0.375) | 2.401 (0.842) |
| **Cholesterol (mg/dL)** | 184.4 (42.6) | 195.8 (37.1) | 213.1 (33.3) | 225.7 (72.8) |
| **Triglycerides (mg/dL)** | 178.9 | 182.1 | 207.4 | 220.4 |
| **History of Stroke (% yes)** | 6.1% | 4.5% | 30.7% | 23.8% |
| **History of MI (% yes)** | 0.0% | 0.0% | 82.7% | 36.6% |
| **Hypertension (% yes)** | 54.6% | 72.7% | 100% | 87.2% |

### 4.1.1 Group A: Patients with both CKD and CVD

The participants from Group A included patients with chronic kidney disease (CKD) and cardiovascular disease (CVD), summing up to a total of 82 individuals. Of this group, 75.6% were male, and the average age was the highest among all groups at 55.59 years (SD = 13.95). The average body mass index (BMI) was also 24.99kg/m² (SD = 3.89), signaling moderate overweight status.

The measure for glycemic control, HbA1c, had an average of 9.14%(SD = 1.92) which indicated poor control. Impairment of renal function was also noted with a mean serum creatinine level of 2.401mg/dL (SD = 0.842) .Total cholesterol was also reported at an average of 225.7mg/dL (SD = 72.8) along with triglycerides at 220.4mg/dL. Cardiovascular comorbidities were also prevalent: 23.8% had a history of stroke, and 36.6% had a prior myocardial infarction (MI). 97.2% of those in the group also suffered from hypertension.

### 4.1.2 Group B: Patients with CKD only

Group B comprised 220 patients with CKD and no history of CVD. This group consisted of 68.2% males, with an average age of 52.64 years (SD = 14.55). Their average BMI was 25.46 kg/m² (SD = 2.61), classifying these patients as overweight.

HbA1c levels indicated reasonable control, averaging 7.80% (SD = 2.39). Signs of renal impairment were present with a mean serum creatinine of 2.274 mg/dL (SD = 0.807). Total cholesterol was, on average, 195.8 mg/dL (SD = 37.1) alongside triglycerides of 182.1 mg/dL and cholesterol levels within the normal range. Stroke history was low at 4.5%, with absent prior MI among participants. Patients with hypertension comprised 72.7% of the group.

**4.1.3 Group C: Patients with CVD only**

Group C consisted of 75 patients with CVD and no evidence of CKD. Male participation was at 34.7%, with average age being 50.59 years (SD = 14.79). Average BMI in this group was 24.95 kg/m² (SD = 5.33).

HbA1c levels reflected poor control of diabetes at 9.22% (SD = 1.79). Serum creatinine was within normal limits at 0.955 mg/dL (SD = 0.375). Total cholesterol was 213.1 mg/dL (SD = 33.3) with triglycerides at 207.4 mg/dL. The patients had a high burden of cardiovascular disease with 30.7% of patients having a stroke and 82.7% had a history of MI. All patients had hypertension.

**4.1.4 Group D: Patients with no CKD and CVD**

Group D included 326 diabetic patients with neither CKD nor CVD. This group comprised 31.0% males and average age in the group was 52.21 years (SD = 10.29). This group also had the highest BMI of 27.41 kg/m² (SD = 4.42).

HbA1c was moderately controlled in this group, as they averaged 8.62% (SD = 2.30). The average serum creatinine value in this group was low at 0.793 mg/dL (SD = 0.483). Total cholesterol and triglycerides were 184.4 mg/dL (SD = 42.6) and 178.9 mg/dL, respectively. This group also had low rates of cardiovascular comorbidities; 6.1% had a history of stroke and no patients reported having MI. Hypertension prevalence was 54.6%.

## 4.2 Statistical Significance and Comparison

To assess the impact of CKD/CVD status on health indicators, one-way ANOVA was conducted on all outcome variables with post hoc analyses performed using both Tukey HSD and Bonferroni tests. Table 4.2 includes an overview of the F-statistics, p-values, and effect sizes ($\eta^2$). Most variables displayed significant differences except for age which had p = 0.082.

*Table 4.2 Summary of one-way ANOVA for each variable (CKD/CVD group as factor).*

| Dependent Variable | F(df=1, 3699) | P | Eta Squared (η²) |
|---|---|---|---|
| Gender (% male) | 40.46 | < 0.001 | 0.148 (large) |
| Age | 2.24 | 0.082 | 0.010 (n.s.) |
| BMI | 17.36 | < 0.001 | 0.069 (moderate) |
| HbA1c | 12.35 | < 0.001 | 0.050 (moderate) |
| Serum Creatinine | 311.40 | < 0.001 | 0.572 (very large) |
| Total Cholesterol | 22.93 | < 0.001 | 0.090 (moderate) |
| Triglycerides | 14.20 | < 0.001 | 0.057 (moderate) |
| History of Stroke | 23.45 | < 0.001 | 0.091 (moderate) |
| History of MI | 402.86 | < 0.001 | 0.634 (very large) |
| Hypertension | 29.14 | < 0.001 | 0.111 (large) |

*Note: F-tests with p < 0.05 are considered significant. Effect size (η²) interpretation is given in parentheses (Cohen's guidelines: 0.01 small, 0.06 medium, 0.14 large). "n.s." = not significant.*

**Group A (Both CKD & CVD):** Significant differences across groups in gender were observed with ANOVA (p < 0.001), with post hoc analyses revealing greater proportions of males in the no-disease and CVD-only cohorts compared. Both BMI and HbA1c differences were significant, particularly with respect to HbA1c being greater than in CKD-only (p < 0.01). Serum creatinine demonstrated a very large effect size and was significantly greater than in both non-CKD groups (p < 0.001). Total cholesterol and triglycerides were also significantly greater than those observed in the CKD-only group (p = 0.002, p = 0.005 respectively) and no-disease (p < 0.001, p < 0.002 respectively). Stroke and MI histories were significantly greater than non-CVD groups (p < 0.001). Hypertension prevalence was significantly greater than the CKD-only (p = 0.018) and no-disease groups (p < 0.001).

**Group B (Only CKD):** We observed significant differences in gender with more males present in the no-disease and CVD-only groups (p < 0.001). With respect to HbA1c, this was significantly lower in comparison to the CVD-only and both-disease groups (p = 0.008 and p < 0.01 respectively). Serum creatinine was found to be much higher compared to all the non-CKD groups (p < 0.001) which is consistent with a CKD diagnosis. Lipid levels were markedly below those in the CVD associated groups. The prevalence of stroke and MI was significantly lower than in the CVD associated groups (p < 0.001). Hypertension was more prevalent than in the no disease group (p < 0.001) but was significantly lower than in both-disease (p = 0.018) and CVD-only groups (p < 0.001).

**Group C (CVD Only):** Significant gender differences were noted where males were fewer in number compared to the CKD-bearing groups. HbA1c was significantly higher than in the group with CKD only (p = 0.008). Serum creatinine was signifiantly lower than in the CKD-bearing groups (p < 0.001). Both Total Cholesterol and Triglycerides were signficantly higer than with the non-CVD groups (p = 0.004 and p = 0.03 vs no-disease; p = 0.03 and p = 0.04 vs CKD-only). The history of Stroke and MI showed highly signficiant differences of which MI was especially high (p < 0.001). Hypertension was significantly higher compared to all other groups (p < 0.001).

**Group D (No CKD or CVD):** Significant changes for this group were noted in lower gender proportion of males, increased BMI compared all other groups (p < 0.001) and having the lowest serum creatinine (p < 0.001 vs both CKD groups). These groups had CVD related events (stroke and MI) significantly lower than in the CVD group (p < 0.01 and p < 0.001 respectively). Hypertension had significantly lessened as compared against every other group (p < 0.001).

These post hoc findings confirm that the presence of CKD and CVD in diabetic patients has a marked impact on several health parameters. The strongest effect sizes were related to serum creatinine and history of MI ($\eta^2$ = 0.572 and 0.634 respectively) which suggests that membership to the group accounts for a great deal of variance in these results. Complete statistical information is depicted in Table 4.2.

## 4.3 Evaluation of Machine Learning Algorithms

This study explains the performance metrics of the three evaluated models Logistic Regression, Support Vector Machine (SVM) and Random Forest with focus on predicting chronic kidney disease (CKD) and cardiovascular disease (CVD) within diabetic population. Each model was trained and tested using stratified cross-validation to provide an unbiased estimate of performance across folds. To mitigate class imbalance (fewer cases of CKD or CVD relative to non-cases), the Synthetic Minority Oversampling Technique (SMOTE) was utilized on the training data for each fold. Classifiers without such balancing are prone to inability to detect minority-class instances because the classifiers become skewed towards the dominant class. SMOTE was beneficial in enhancing the models' sensitivity to the minority outcomes (CKD or CVD) by creating synthetic instances of these outcomes during the training process. The results are structured by model: first the performance of Logistic Regression on CKD and CVD prediction tasks, then SVM, and finally Random Forest were described. Significant classification benchmarks – AUC, accuracy, precision, recall, F1-score are provided for each model. Summary classification metrics for each model are displayed in Table 4.3 and Table 4.4 for the CKD and CVD prediction tasks over cross validation and Table 4.5 and 4.6 for the 80:20 train test split prediction, respectively. Furthermore, Table 4.7 and 4.8 shows the accuracies of different models on different stratified cross validation folds for CKD and CVD prediction.

*Table 4.3 Classification performance metrics for CKD prediction by each model (averaged over cross-validation).*

| Model | AUC | Accuracy | Precision | Recall | F1-score |
|---|---|---|---|---|---|
| **Logistic Regression** | 0.78 | 79.0% | 48.8% | 69.0% | 57.1% |
| **SVM** | 0.82 | 76.9% | 45.0% | 62.1% | 52.2% |
| **Random Forest** | 0.98 | 95.8% | 87.1% | 93.1% | 90.0% |

*Table 4.4 Classification performance metrics for CVD prediction by each model (averaged over cross-validation).*

| Model | AUC | Accuracy | Precision | Recall | F1-score |
|---|---|---|---|---|---|
| **Logistic Reg.** | 0.94 | 90.9% | 46.7% | 58.3% | 51.9% |
| **SVM** | 0.95 | 93.0% | 60.0% | 50.0% | 54.5% |
| **Random Forest** | 0.91 | 90.9% | 47.6% | 83.3% | 60.6% |

*Table 4.5 Classification performance metrics for CVD prediction by each mode (train test 80:20 split)*

| Model | AUC | Accuracy | Precision | Recall | F1-score |
|---|---|---|---|---|---|
| **Logistic Reg.** | 0.93 | 89.7% | 33% | 50% | 40% |
| **SVM** | 0.96 | 89.7% | 33% | 50% | 40% |
| **Random Forest** | 0.98 | 93.1% | 50% | 50% | 50% |

*Table 4.6 Classification performance metrics for CKD prediction by each mode (train test 80:20 split)*

| Model | AUC | Accuracy | Precision | Recall | F1-score |
|---|---|---|---|---|---|
| **Logistic Reg.** | 0.98 | 89.7% | 80% | 66.7% | 72.7% |
| **SVM** | 0.97 | 82.8% | 100% | 16.7% | 28.6% |
| **Random Forest** | 0.94 | 86.2% | 75% | 50% | 60% |

*Table 4.7 Accuracy CKD Table (Stratified CV from 5 to 9 folds)*

| Model | Stratified Cross-Validation Folds | | | | |
|---|---|---|---|---|---|
|  | **5** | **6** | **7** | **8** | **9** |
| **Logistic Reg.** | 78% | 79% | 79% | 76% | 75% |
| **SVM** | 76% | 77% | 77% | 80% | 78% |
| **Random Forest** | 95% | 95% | 96% | 94% | 94% |

*Table 4.8 Accuracy CVD Table (Stratified CV from 5 to 9 folds)*

| Model | Stratified Cross-Validation Folds | | | | |
|---|---|---|---|---|---|
|  | **5** | **6** | **7** | **8** | **9** |
| **Logistic Reg.** | 93% | 94% | 91% | 94% | 92% |
| **SVM** | 92% | 92% | 93% | 93% | 93% |
| **Random Forest** | 93% | 92% | 91% | 90% | 91% |

## 4.3.1 Logistic Regression

### 4.3.1.1 CKD Prediction

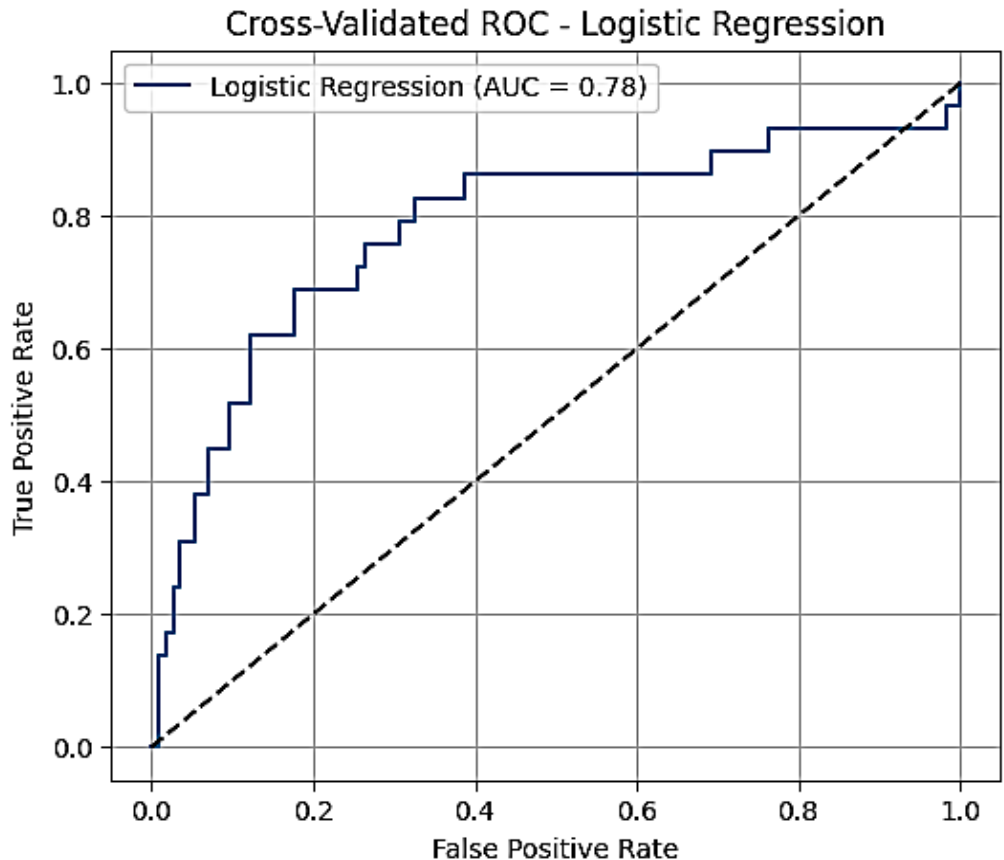

*Figure 4.1 Cross-validated ROC curve for LR (CKD)*

The ROC curve reflects CKD prediction sensitivity and specificity trade-offs, as well as model accuracy which ROC verifies is greater than random chance. An AUC of 0.78 is moderate in distinguishing diabetic with CKD from those without. It is widely accepted that with AUC values between 0.7 to 0.8 moderate discrimination occurs. The regression model attained 79% accuracy on the CKD prediction task (Table 4.3). However, the precision and recall metrics demonstrate a difference in performance: precision at 48.8% and recall at 69.0% (Table 4.3) suggest relatively lower positive predictive value despite reasonable sensitivity. In simpler terms, when utilizing regression for CKD diagnosis, it can be said that about one in two patients marked by the model as "CKD positive" indeed had CKD but about two-thirds of all actual CKD cases were identified correctly. The F1 score of 0.57 means CKD was modestly defended by the blend of precision and recall counterbalancing each other suggesting performance on the minority class was dull.

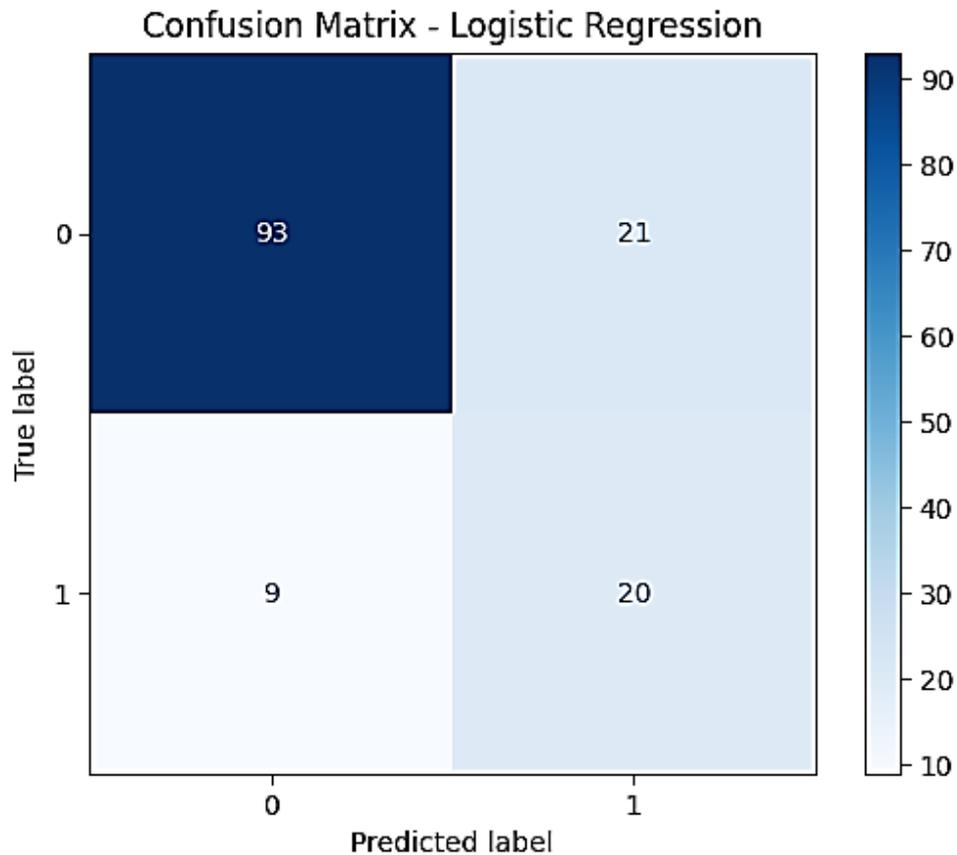

*Figure 4.2 Confusion matrix LR (CKD)*

The confusion matrix (Figure 4.2) provides additional details regarding classification results. In the evaluation (which combines all cross-validation folds), out of 143 total instances, the model achieved true positive identification of 20 out of 29 CKD cases and true negative identification of 93 out of 114 non-CKD cases. Nevertheless, 9 CKD patients were misclassified as non-CKD (false negatives), and 21 non-CKD patients were misclassified as having CKD (false positives). These findings are consistent with the previously mentioned recall (69% detection of CKD cases) and precision (49% estimation of CKD cases was confirmed). From a clinical standpoint, the false negatives (9 missed CKD cases) are troubling because those patients may not be referred for timely evaluation and intervention by a nephrologist. The false positive figure (21 non-CKD patients labeled as having CKD) suggests that some patients would be subjected to unnecessary follow-up tests. Although these claims can cause patient anxiety and experience costs to the healthcare system, in this situation, they are less harmful than the overwhelming impact of false positives—without timely management, progression of CKD could occur undiagnosed.

The application of SMOTE during the training process likely aided the model in identifying the majority of CKD cases. Without oversampling the minority class, logistic regression would have missed additional CKD cases because of its bias toward the prevalent class. Cross-validation further ensured that the reported performance is reproducible across multiple data partitions, mitigating concerns that the results might stem from encountering a single fortunate or unfortunate divide.

**4.3.1.2 CVD Prediction**

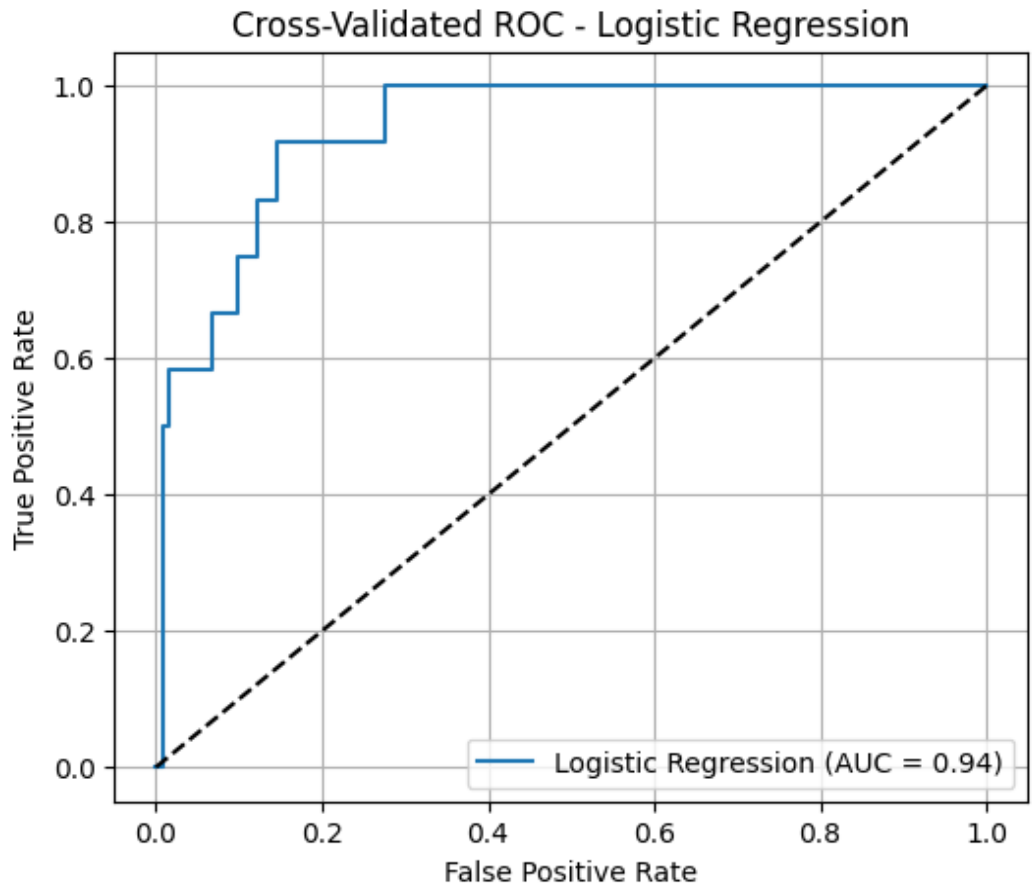

*Figure 4.3 Cross-validated ROC curve for LR (CVD)*

The Logistic Regression model performed vastly better on CVD prediction than CKD, which is apparent from the steep ROC curve in Figure 4.3. The model achieved 0.94 AUC which reflects superb discriminative ability (AUC > 0.9 is generally considered outstanding). This high AUC means that irrespective of the classification cutoff point, the logistic model was able to predict accurately the rank order of the diabetic patients with CVD and those without CVD (i.e. patients who had CVD were assigned higher prediction scores compared to those who did not). CVD prediction accuracy overall was 90.9% (Table 4.4), again yielding better results than CKD indicating stronger model performance on this task. For the CVD class, precision was 46.7% and recall was 58.3% (Table 4.4). Interestingly, the CVD recall figure (58.3%) is lower than the CKD figure, so the CVD gratefulness outrank CA moderate CVD possess a relative degree CVD impairment recognition despite higher AUC submission. This observation suggests that the model's ability to accurately classify patients as high or low risk of CVD was very good given the high AUC; however, at the default threshold most predictions were set too low and class imbalance with choice of threshold likely drove this effect.

The accuracy rate (46.7%) aligns with CKD's accuracy, which indicates lower than half of the positive diagnosis were actual CVD cases, which demonstrates the infrequency of CVD outcomes in the population and that the model is biased toward under-diagnosing.

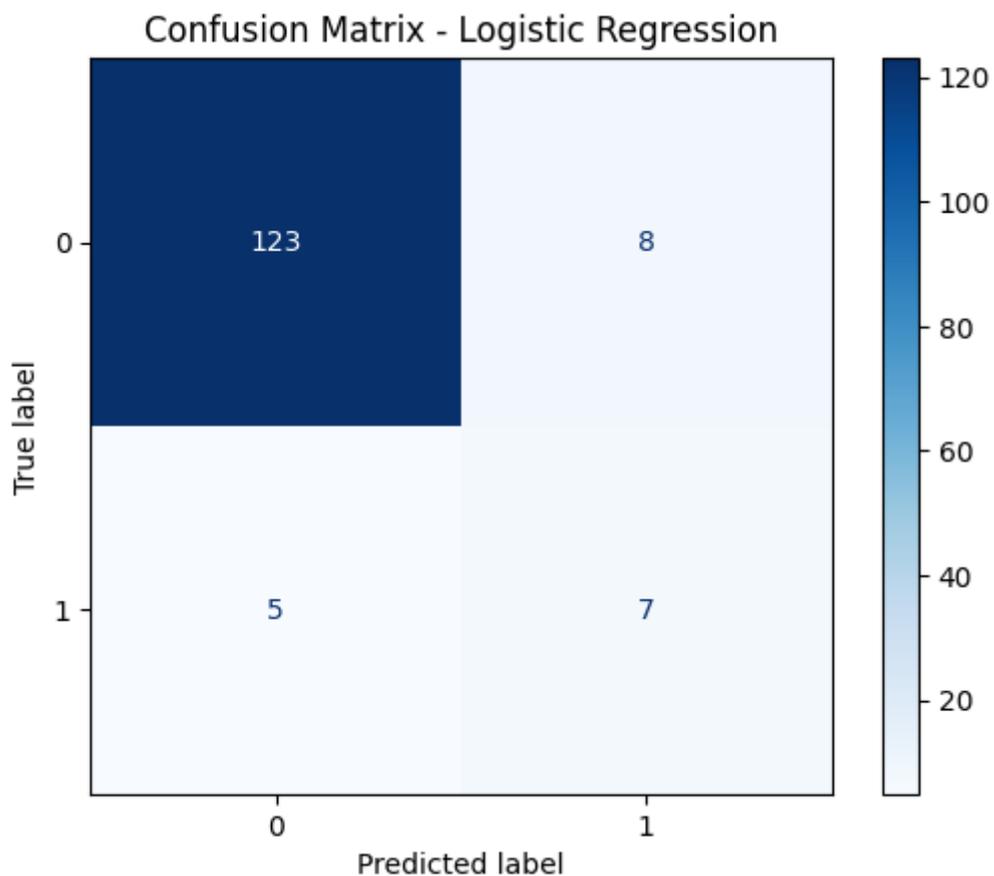

*Figure 4.4 Confusion matrix for LR (CVD)*

The confusion matrix (Figure 4.4) indicates that logistic regression successfully recognized 7 of 12 patients with CVD (true positives) and 123 out of 131 patients without CVD (true negatives). The model missed capturing 5 CVD patients (false negatives) and inaccurately labeled 8 non-CVD patients as having the disease (false positives). The predominant true negatives speaks to an effort to balance sensitivity and specificity - the model avoided misclassifying the majority of non-CVD individuals.

Yet, failing to identify 5 CVD cases (about 41.7% of the actual CVD cases) suggests a large number of undetected true positives at the set threshold. From a clinical standpoint, those cases would be quite troubling because undiagnosed cardiovascular risk could mean a lack of needed preventive treatment. The false positives (8 patients) would mean some patients lacking CVD would be subjected to unnecessary cardiac evaluations or treatments.

In comparison with CKD results, logistic regression demonstrated greater specificity for CVD (fewer false positive results) but lower sensitivity (also known as recall). This result might be affected by the much smaller portion of positive cases in the CVD dataset (12 out of 143 as opposed to 29 for CKD), which could lead to overly aggressive negative class bias shift in threshold calibration. The high cross-validated AUC of 0.94 suggests that altering the classification threshold could achieve a more optimal sensitivity specificity balance, e.g. setting the threshold slightly lower would allow the model to better than 7 CVD diagnoses while still maintaining acceptable specificity due to the strong ROC curve. In CVD, as opposed to CKD, the logistic regression performed optimally, indicating that this cohort was more responsive to the model's attempt at predicting cardiovascular outcomes.

## 4.3.2 Support Vector Machine (SVM)

### 4.3.2.1 CKD Prediction

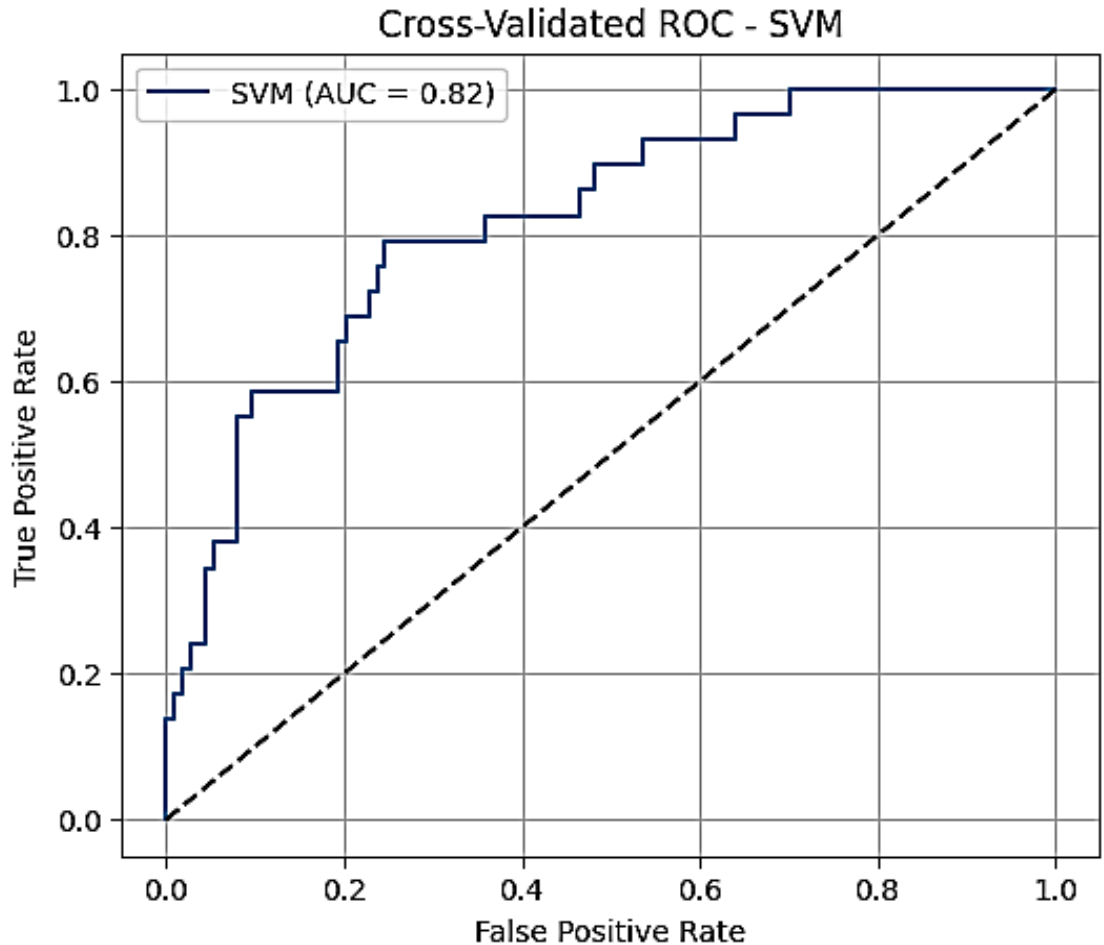

*Figure 4.5 Cross-validated ROC curve for SVM (CKD)*

The SVM model with a kernel-based decision boundary demonstrated better discriminative power for CKD AUC compared to logistic regression with 0.82 (Table 4.3). An AUC in this range is considered good, indicating the model was better able to stratify patients by their risk for CKD. The ROC curve in Figure 5 rises steeper than the logistic curve, meaning the SVM outperformed the logistic regression model for rate of true positives achieved for a given rate of false positives. The cross-validated accuracy of SVM on CKD was 76.9%, which is lower than that of logistic regression 79.0% (see Table 4.3). More strikingly, the precision for CKD was 45.0% and recall 62.1% (Table 4.3). From these values, it can be estimated that out of all the patients the SVM predicted to have CKD, 45% truly had the disease, and the model detected about 62% of the total CKD cases.

When compared to logistic regression, SVM's recall performance for CKD was somewhat lower. Recall was 62.1% versus 69.0% and precision was 45.0% versus 48.8%. This indicates that the SVM, at the default threshold, was more conservative in determining CKD positives, or put another way, it missed more CKD cases while still generating approximately the same number of false positives as the logistic regression. SVM's F1 score on CKD was 0.52 and logistic regression's was 0.57 and both are very close, which strengthens the notion that the predictive power of the CKD class doesn't differ significantly. Notice the higher AUC from the SVM indicates that with a more optimal threshold or cost ratio, it could be more accurate than the logistic regressor in CKD case detection. The model has a ROC-optimally-driven decision function which excels in class separation, despite yielding similar confusion matrix results to logistic regression due to the threshold chosen. Allowing SVMs to utilize SMOTE during their training was critical for providing sufficient examples from the minority class; otherwise, SVMs tend to bias toward the overly represented class in cases of extreme imbalance within training data. With SMOTE, SVM's decision boundary for CKD was more inclusive of positive cases, leading to the decent recall that was observed.

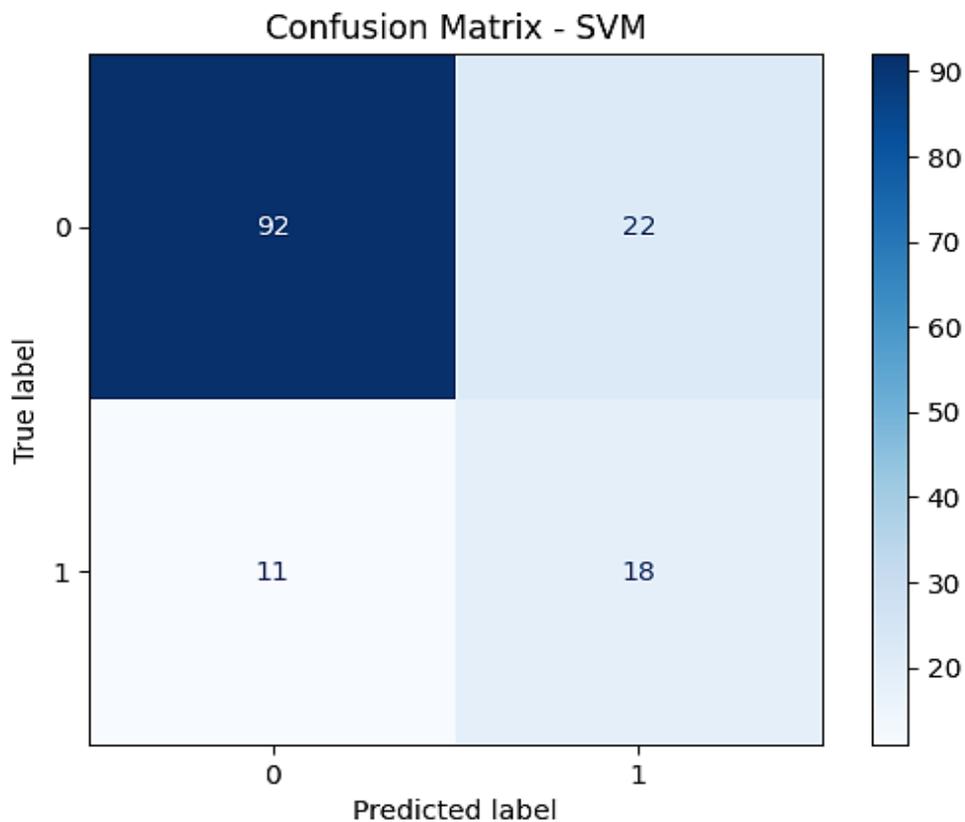

*Figure 4.6 Confusion matrix for SVM (CKD)*

According to Figure 4.6, the SVM correctly identified 18 of the 29 CKD patients and 92 of the 114 non-CKD patients correctly. This yielded 11 false negatives (missed CKD patients) and 22 false positives (non-CKD patients incorrectly identified as CKD). These numbers are consistent with the precision and recall: the SVM missed slightly more cases than the regression model (SVM had 11 FN vs. logistic's 9 FN), which is why recall (62.1%) is lower. It also generated approximately the same number of false positives (22 for SVM and 21 for logistic) accuracy just lower than before. In the SVM false negatives clinically (11) means out of 38% actual CKD cases would go undetected which is concerning for patient care as those individuals might receive early intervention. The false positive (22) would add collapse of described for logistic regression – additional tests for some patients without CKD.

The current balance implies that the SVM, similar to logistic regression, encountered difficulties with some CKD cases because those patients likely feature overlaps with non-CKD patients, leading to those omissions. Training with SMOTE augmentation helped the SVM obtain 18 CKD cases; without oversampling, true positive count could have been lower since SVM may have draped the majority class. In total, regarding CKD prediction, SVM showed a moderate increase in AUC over logistic regression outweighing AUC of 0.82 and 0.78 respectively, which suggests in subsantially exceeding model capability, though not significantly increasing recall or precision at the fixed threshold with 0.78. These findings indicate the SVM's edge might be in flexible thresholds for example, if tuning the decision function cut-off were needed to improve sensitivity, it could certainly be done.

**4.3.2.2 CVD Prediction**

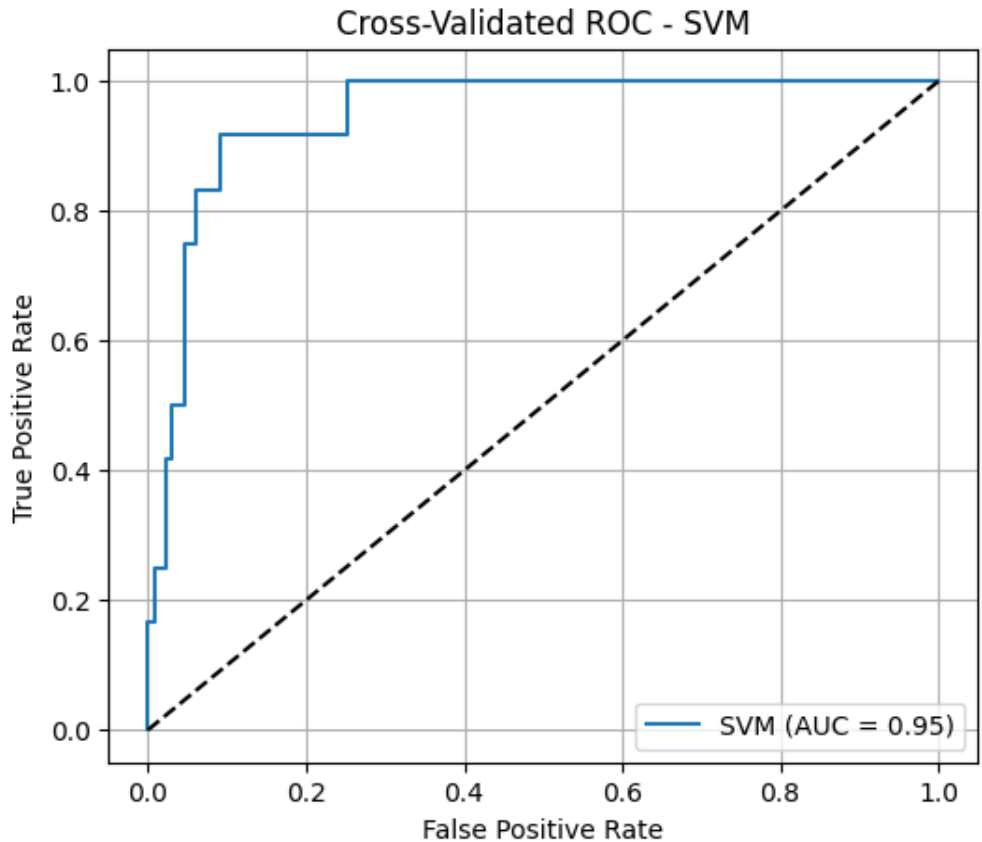

*Figure 4.7 Cross-validated ROC curve for SVM (CVD)*

The SVM model achieved the highest AUC (0.95) among the three models for CVD prediction (Table 4.4). This means that the model had excellent discriminative power. Furthermore, the ROC curve for SVM on CVD shown in Figure 4.7 is steep and located at the upper left corner, meaning that the model has high true positive rates while simultaneously maintaining low false positive rates. An AUC of 0.95 is excellent and suggests that the SVM was very efficient in the ranking of diabetic patients in relation to the likelihood of having CVD. The model's cross-validated accuracy reached approximately 93.0%. This is the highest observed for the CVD models in table 4.4. Despite these strong results, the metrics derived from the confusion matrix indicate an interesting balance. The SVM's precision for CVD was 60% much better than that of the logistic regression while the recall figure was 50% which was the lowest of the three models on CVD (Table 4.4). To put it another way, the SVM predicted a patient with CVD correctly 60% of the time which means when the SVM predicted there was a patient with CVD, he was right 60% of the time showing there were few patients falsely labelled as such. However, it only detected half of the actual CVD cases.

The F1-score obtained for SVM on CVD was 0.545, aligning with the other models' CVD F1-scores which were within the range of 0.52–0.61. The SVM's behavior strongly suggests that a sharper decision boundary was set for CVD due to class imbalance (only 12 positive cases). This fits the high AUC well: the model gives high probability estimates to true CVD cases (allowing excellent ROC), but at the default threshold of 0.5, it classified fewer patients as positive to reduce false alerts. Increasing the threshold could improve recall while lowering precision, which as stated, the model did prefer. The incorporation of SMOTE when training the SVM for CVD was important because the positive class was extremely small; oversampling guaranteed that the SVM had enough synthetic CVD cases to learn from. Regardless of the conservative evaluation cut point and high AUC sustaining the SVM knowing the distinguishing data patterns, the patterns were learned quite well.

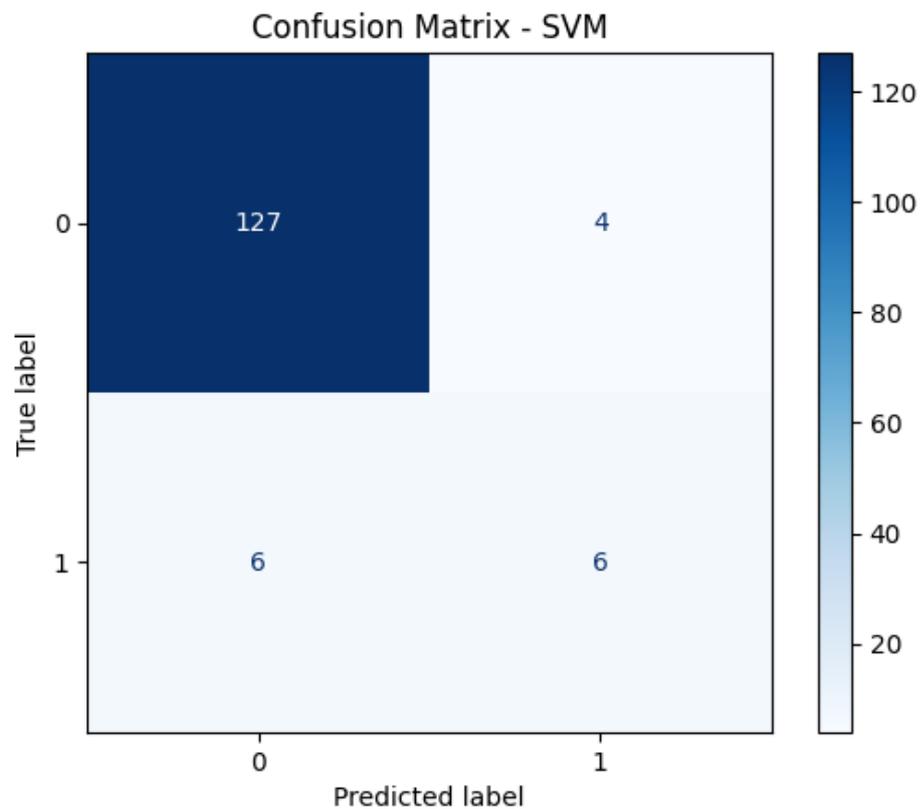

*Figure 4.8 Confusion matrix for SVM ( CVD )*

The confusion matrix in Figure 4.8 illustrates SVM's under predictive nature for CVD. The model diagnosed 6 out of 12 CVD cases correctly (TP) and 127 out of 131 non-CVD cases (TN) correctly. This produced 6 false negatives (missing half the CVD cases) and only 4 false positives. The false positive rate here is extremely low (4 out of 131 non-CVD = ~3%) which supports the high precision (60%) of the model. SVM in this case means fewer unnecessary interventions (only 4 people would be falsely categorized as high risk for CVD), but more missed patients (6 high risk CVD patients not identified). If this model was used for screening, it might be good for situations where one wants to avoid raising false alarms for example, if the subsequent test for CVD risk is costly or invasive, then precision (referring less people fraudulently) would be more desirable. From a clinical point, SVM shifted the burden of intervention to patients which translates to fewer automated alerts tuned to annoy providers; instead, they redirect care to patients who need it. However, the large issue is flagging half the actual cases. Those patients might go without proactive cardiology care and face bad outcomes.

The missed cases were likely the ones with some lower SVM output scores. If the threshold were lowered, many of those could be caught, which underscores the flexibility offered by the SVM ranking performance. As previously noted, the SVM model's CVD results exhibit strong accuracy and discriminative power alongside a precise-detection-versus-recall imbalance; the model's specificity resulted in low false positives but at the expense of sensitivity. This precisely demonstrates class imbalance and model calibration influence - rebalancing with SMOTE did not offset the cautious assumption CVD case rarity imposed on the SVM's prediction strategy.

### 4.3.3 Random Forest

#### 4.3.3.1 CKD Prediction

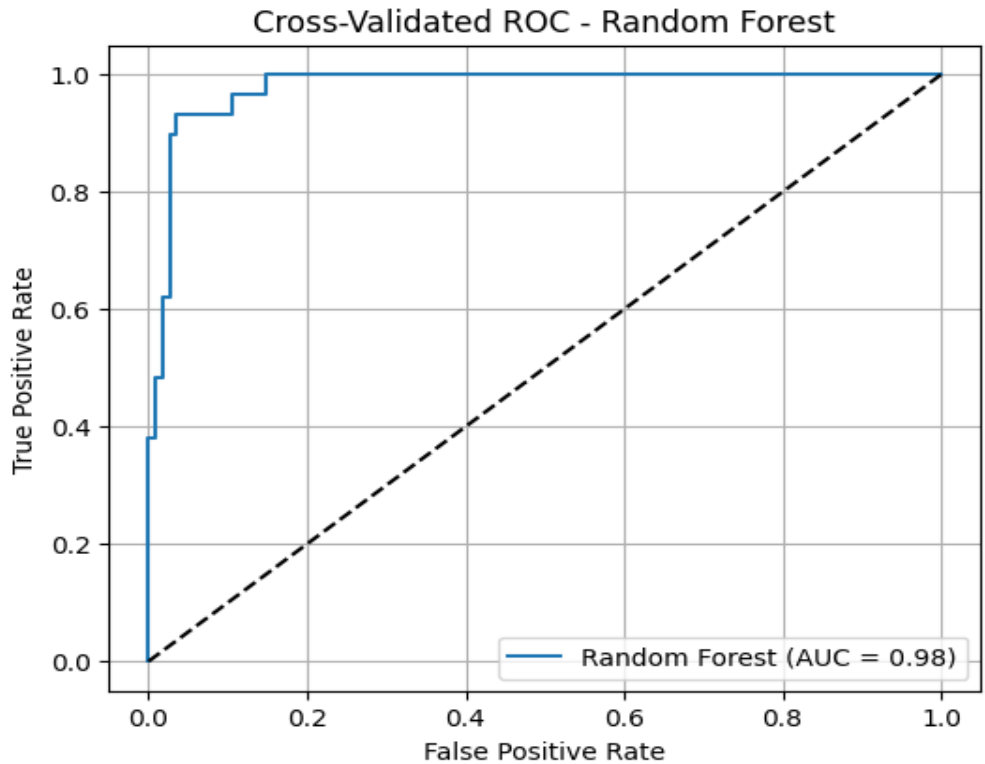

*Figure 4.9 Cross-validated ROC curve for RF (CKD)*

The Random Forest achieved outstanding performance on the CKD prediction task. Its ROC curve (Figure 4.9) is almost touching the top-left corner of the plot, and the AUC is 0.98, which is in the excellent range and indeed close to a perfect score. This indicates the ensemble model was able to almost perfectly discriminate between CKD and non-CKD patients in the dataset. The cross-validated accuracy was 95.8% (Table 4.3), substantially higher than that of logistic regression or SVM for CKD. Such high accuracy reflects the model's ability to get most predictions correct. The precision of Random Forest on CKD was 87.1%, and recall was 93.1% (Table 4.3), which are extremely high for a minority-class prediction problem. In fact, these values suggest that the Random Forest correctly identified the vast majority of CKD cases while also keeping false positives very low. The F1 score was 0.90, indicating a very well-balanced high precision and recall.

These findings showing Random Forest model dominating the other two models regarding CKD prediction in this particular diabetic cohort indicates its nonlinear feature interactions and multidimensional interrelations capabilities (which logistic regression would miss as a linear model). Ensemble averaging also contributed to the overfitting robustness. Employing SMOTE further ensured positive CKD-exemplar exposure for learning during training, enabling subtle CKD detection. Cross-validation supports consistent performance confirming lack of fluke with partition dependence. Based model usage explanatory classification guidelines AUC near 0.98, essentially interpreting it as definitive borderless test, while no model exists without imperfections, evidence indicates Random Forest almost flawlessly delineated CKD and non-CKD patients with incredible accuracy, minimal error.

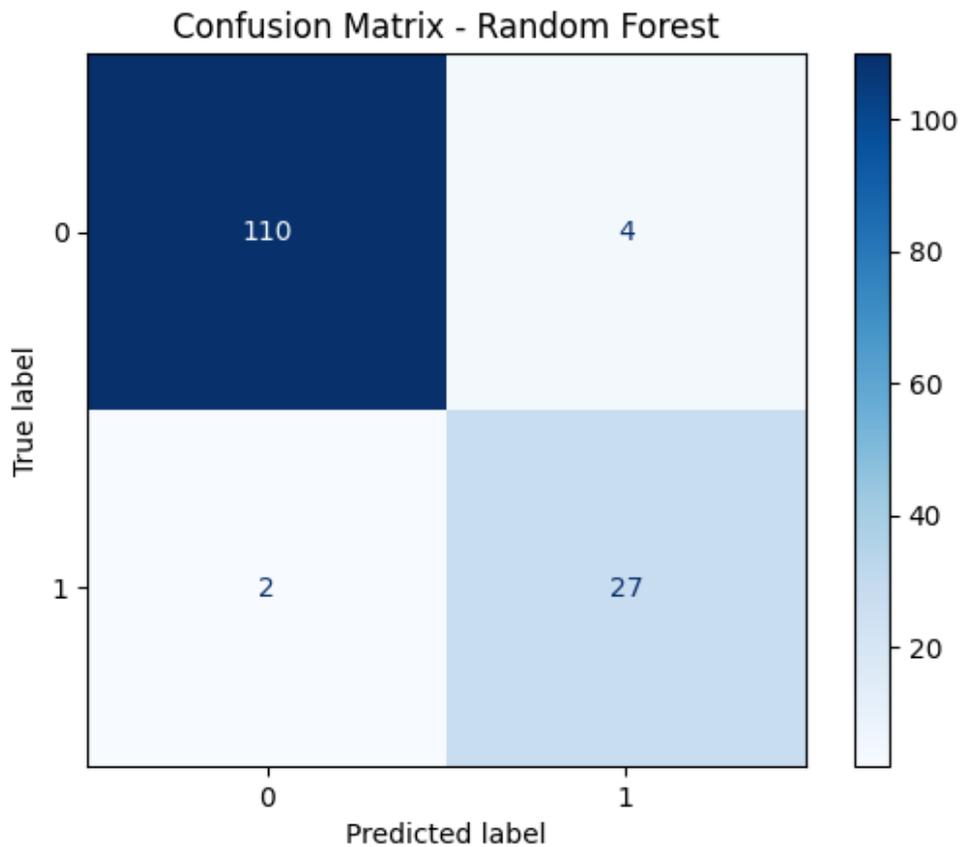

*Figure 4.10 Confusion martrix for RF (CKD)*

The Random Forest's astonishing prowess on CKD is corroborated by the confusion matrix (Figure 4.10). Of the 29 actual CKD patients, the model successfully recognized 27 as CKD. This means missing only 2 cases (false negatives). Out of 114 non-CKD patients, the model accurately classifed 110 of them as non-CKD (true negatives) and incorrectly classified 4 as non-CKD. Put differently, the model attained 93.1% sensitivity (recall) and 96.5% specificity for CKD, which is an exquisite balance. The false negatives = 2 indicates that only a very small percentage of CKD cases would be assumed to be missed. Clinically, this translates to the vast majority of patients with CKD in the dataset would be model flagged, ensuring follow-up care. The false positives = 4 implies that only those 4 individuals without CKD, would be incorrectly assumed to be at risk and would be rendered as truly nominal or microscopically non-significant (rounding supposedly 3.5% of 114 non-CKD). While such patients may be subjected to needless procedures, the burden remains minimal because the ROI of capturing nearly all true cases is so high.

The Random Forest confusion matrix reveals its high precision of 87% (far greater than the 45-49% precision from logistic and SVM) meaning far more precisely predictive. This precision is extraordinarily useful in a clinical screening context because the model's alerts are mostly valid. The high recall (93%) is equally useful because it reduces the number of missed diagnoses. The ability of Random Forest to achieve both high precision and recall on the CKD problem with class imbalance is very likely an indication of thorough model tuning combined with the strength for ensemble learning and data balancing using SMOTE. Most noteworthy, however, such attribution could also suggest an aspect of overfitting to the cross-validation folds, but performing cross-validation and observing consistent results across folds (as evidenced by stable ROC and metrics) supports the model's performance being reliable. To conclude, Random Forest with this dataset achieved an almost ideal CKD predictor and surpassed all other models to this aid.

**4.3.3.2 CVD Prediction**

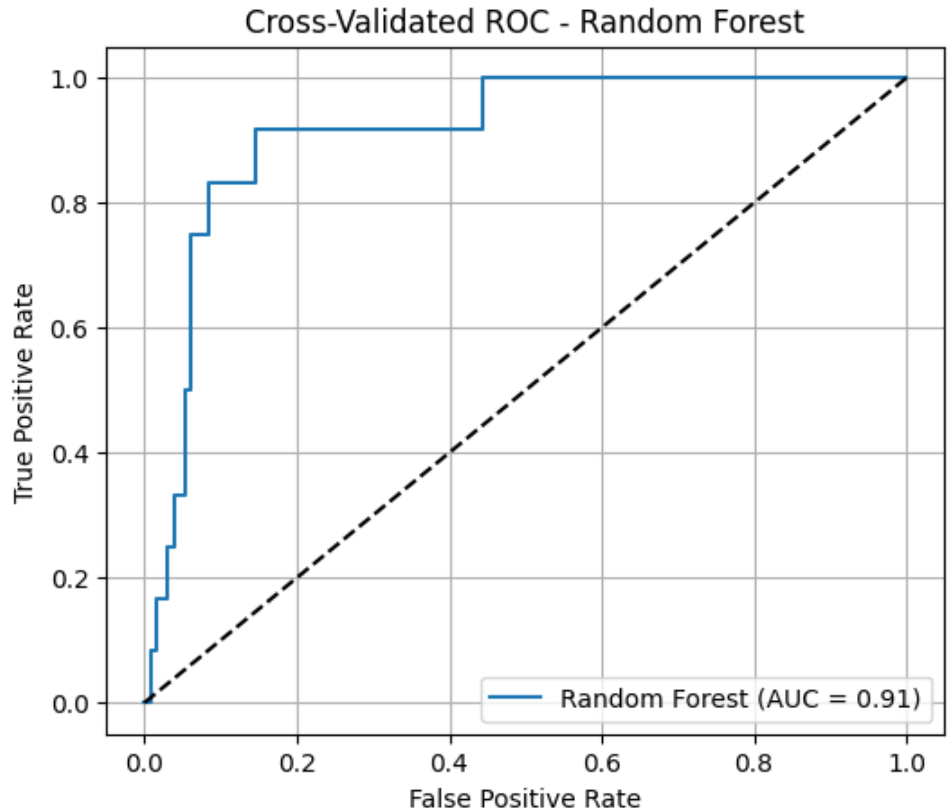

*Figure .4.11 Cross-validated for RF (CVD)*

Random Forest retained its predictive prowess on the CVD prediction task, but its margin of superiority relative to the other models was not as pronounced as in CKD. As shown in Figure 4.11, the ROC curve for Random Forest CVD prediction displays AUC of 0.91, which indicates excellent discrimination overall, but notably this AUC is now lower than the logistic regression and SVM scores of 0.94 and 0.95 respectively. Although still well above the diagonal, CVD 90.9% accurate classification with Random Forest suggests a reasonable level of model discrimination between CVD and non-CVD patients, but not SVM peaky elevation on the ROC curve. In Table 4.3, Random Forest's accuracy for CVD was 90.9%, essentially tied with logistic regression and some number below SVM's 93.0%. Precision for CVD was 47.6% and recall achieved was 83.3%, resulting in an F1 score of 0.606, the highest among the three models for CVD. These measurements indicate that unlike SVM, with Random Forest the tradeoff was tilted toward better recall for CVD.

Particularly the Random Forest captured a large majority of CVD cases (recall 83% is much higher than logistic's 58% or SVM's 50%) While its precision of 47.6% was lower than that of SVM 60% and comparable to logistic 46.7%. This tells us that there is more false positive instead of fewer false negative results. The approach reflected here is typical for a model that tries not to miss true cases. It is likely that the Random Forest had a lower internal threshold for classifying someone as CVD positive, or due to ensemble voting, it produced stronger signals for positives. Recall is high in many medical contexts (because missing a CVD case is dangerous), but the downside is substantial: almost half of the positive predictions were wrong. Random Forest's ranking of patients by CVD risk is strong, but not quite as flawless as it was for CKD deducts a 0.91 AUC ROC, which tells us there is likely more overlap in the feature patterns of CVD vs non-CVD within the data. This is one reason all models, including Random Forest, did not attain the near perfect AUC seen in CKD. Regardless, an AUC above 0.90 still signifies outstanding delineation.

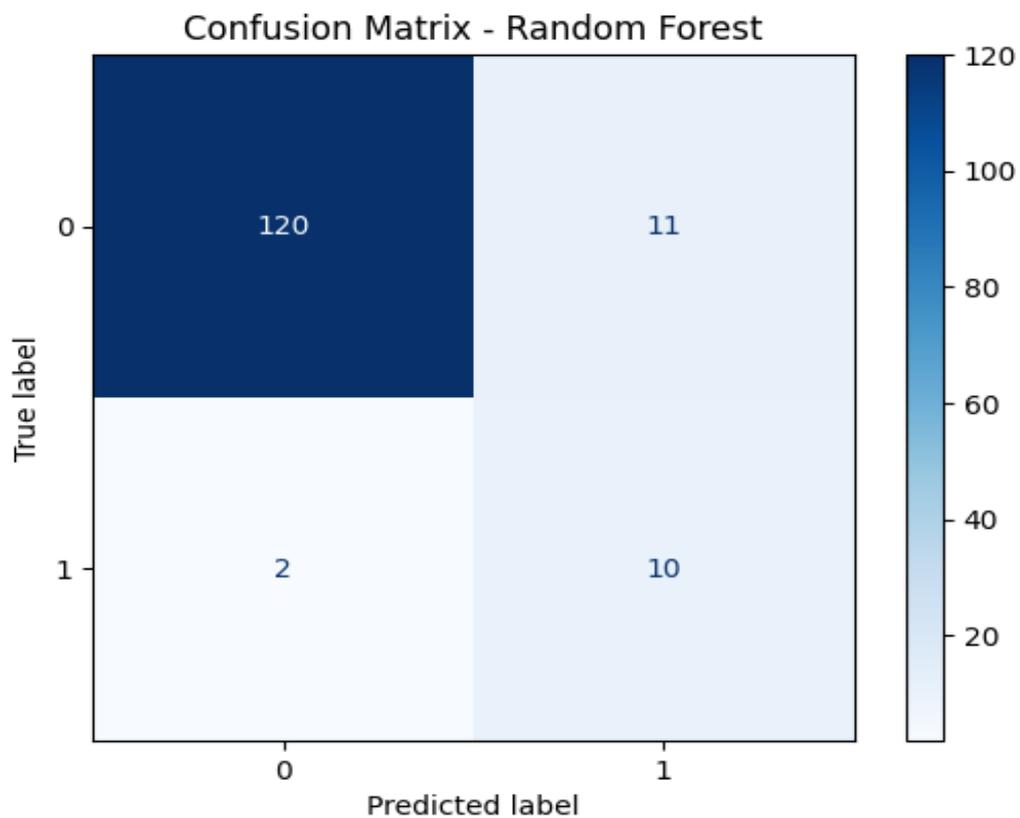

*Figure 4.12 Confusion Matrix for RF (CVD)*

The confusion matrix generated for Random Forest on CVD (Figure 4.12) indicates that it correctly recognized 10 out of 12 CVD cases (true positives=10) and correctly identified 120 out of 131 as non-CVD patients (true negatives=120). There were 2 missed CVD cases (false negatives) and 11 patients flagged as false CVD cases (false positives). The outcome underscores the model's greater focus on sensitivity: a vast majority of cases (83.3% recall) were not missed—significantly high for a model. The tradeoff was greater false alarms. Sensitivity outperformed specificity, evidenced by lower precision. Clinically, those numbers mean nearly all patients who actually have CVD risk would be detected—only 16.7% of CVD cases evaded detection—which bolsters the argument in favor of using the model as a screening or early alert system for cardiovascular concerns. The missed maybe had idiosyncratic patterns beyond the model's generalization, but that's still a small miss rate.

On the bright side, 11 false positives mean that some patients would be subjected to unnecessary follow-ups for CVD risk—perhaps optional diet or lifestyle interventions, as well as medical testing that would otherwise be unneeded. This trade-off may or may not be acceptable, depending on the situation. In preventive cardiology, false positives are preferable in exchange for heightened sensitivity, particularly when the screening comfort measures (lifestyle recommendations or non-invasive testing) are benign. The performance of Random Forest can be compared with that of SVM's on CVD: RF had 10 TP while SVM had 6 TP (thus saving four additional CVD patients), and 11 FP while SVM had 4 FP (seven additional false alerts). Logistic regression was in between being quite moderate (7 TP, 8 FP). These differences demonstrate how greatly Random Forest seems to have favored an overly optimistic prediction policy. SMOTE use in training assisted Random Forest in not overshooting the underrepresented class; it was able to take most CVD signals and positively confirmed cases.

Other than that, the ensemble characteristic of Random Forest (which aggregates many decision trees) serves to strengthen it against noise and allows it to model complex interactions among features, which likely helped it achieve those 10 true positives. The Random Forest model, despite having a slightly lower AUC compared to SVM, showcased better classification balance (evidenced by the highest F1-score for CVD) and practical usability in CVD detection. Therefore, amongst other models, Random Forest provided the most sensitive CVD predictions, ensuring the lowest possible false-negatives with a reasonable count of false-positives. Throughout both outcomes of the disease, Random Forest stood out as the exceptional model for case detection (especially for CKD), demonstrating the effectiveness of ensemble approaches with class imbalance mitigation in clinical prediction scenarios.

## 4.4 Comparative analysis (Train Test Split and Cross Validation)

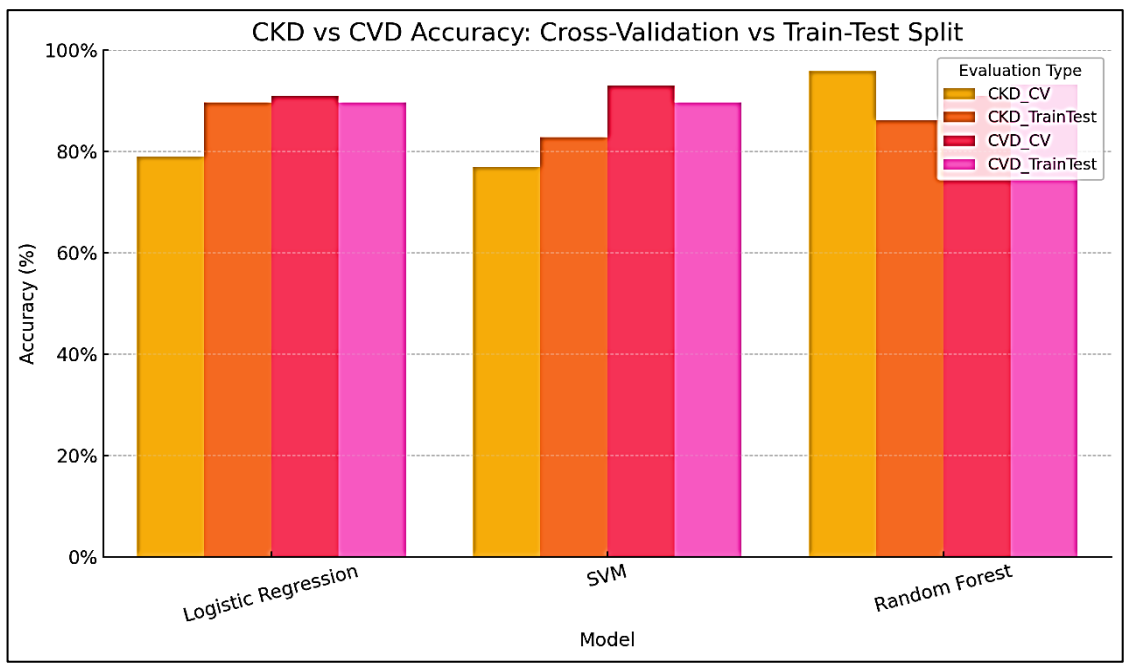

*Figure 4.13 Accuracy comparision of CV and Train-Test Split methods*

In Figure 4.13, model accuracy comparisons for Cross Validated and Train-Test Split methods for Chronic Kidney Disease (CKD) and Cardiovascular Disease (CVD) prediction are presented. The comparison includes three commonly used machine learning models: Logistic Regression, Support Vector Machine (SVM), and Random Forest. In both of the diseases, the Random Forest model achieved the best performance by dominating over other models which demonstrates its high capability in identifying complex relationships in the data. Logistic Regression and SVM showed lower accuracy with the worst performance during CKD prediction. Also, cross validation tends to be more stable and reliable in estimating performance than train-test-split, particularly when overfitting is prevalent. These findings reinforce the effectiveness of ensemble models and multi-factor evaluation methods for clinical predictive modeling challenges.

## 4.5 Random Forest Feature Importance

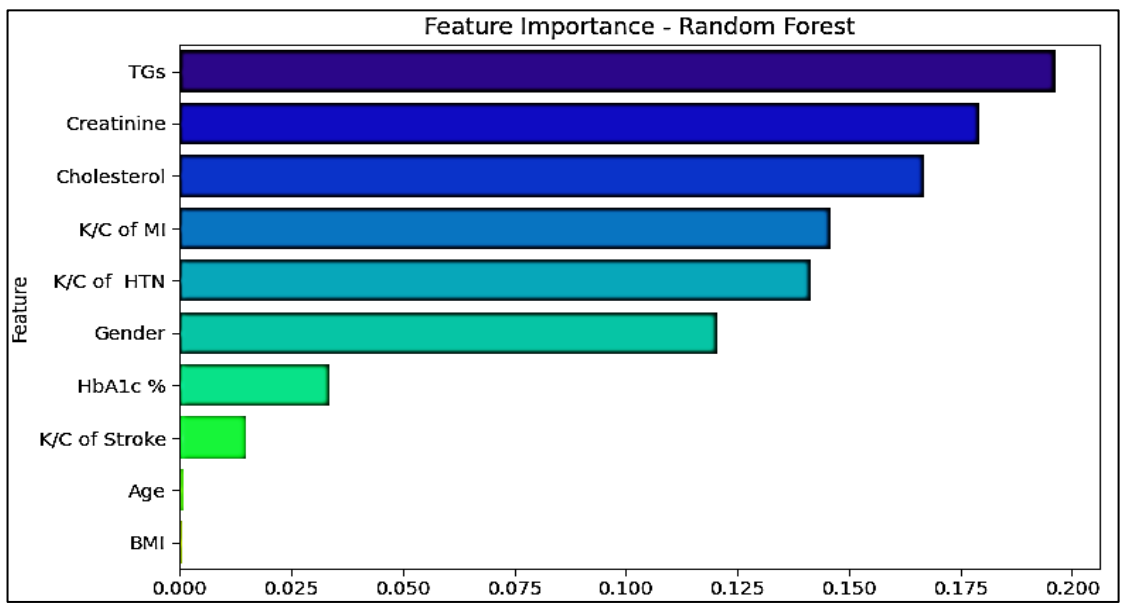

*Figure 4.14 RF Feature importance in CVD prediction*

Figure 4.14 shows the Random Forest Model with cardiovascular disease (CVD) predictors for diabetic patients alongside the extracted feature importance scores. The model extracts triglycerides (TGs), creatinine, and cholesterol as the strongest predictors which highlights the importance of the lipid profile as well as renal function to evaluate the cardiovascular risk in this population. From the clinical history, Known case of myocardial infarction (K/C of MI) and Hypertension (HTN) are also noted to be very significant which reflect their well-known impact on CVD. Gender as well as markers of long-term glycemic control like HbA1c% have moderate importance which indicates that these factors are meaningful to risk estimation. On the contrary, age, history of stroke, and BMI were given less importance in the model which could have been due to low variance and indirection in the dataset these factors have. This perspective perfectly fits the clinical anticipations and provides more trust to the model for ranking clinically relevant features, which increases the capability of the model to be used in decision-support systems for early CVD prediction in diabetic patients.

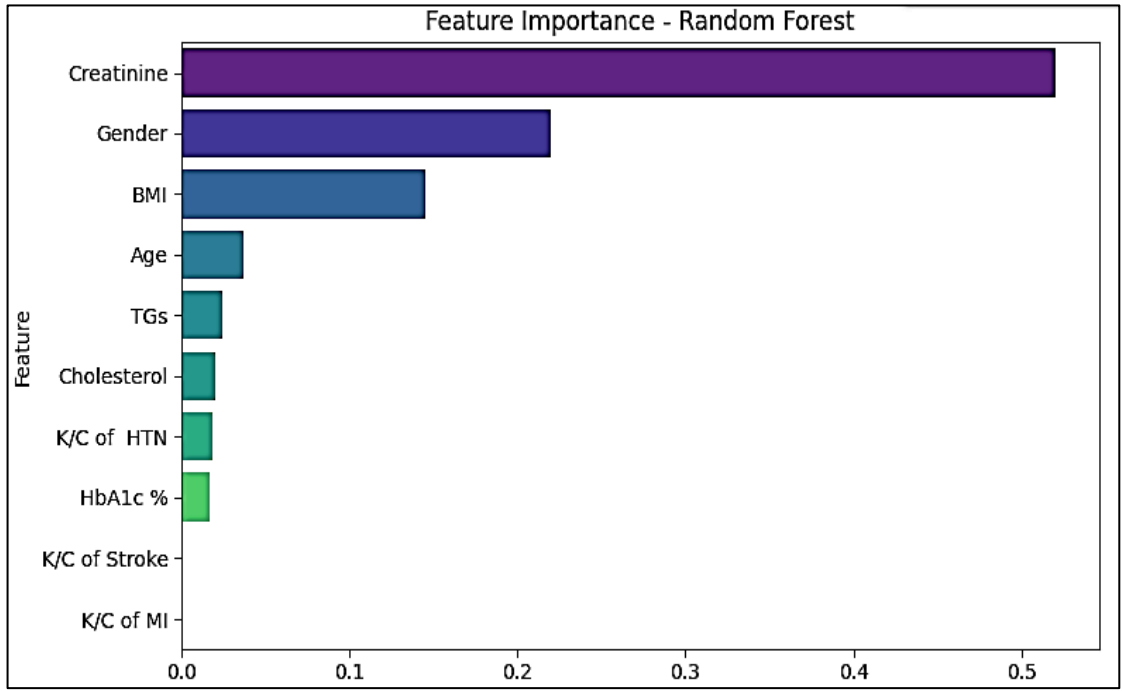

*Figure 4.15 RF feature importance in CKD prediction*

Figure 4.15 illustrates the key features identified by the Random Forest model predicting Chronic Kidney Disease (CKD) in patients with diabetes. Among all the input features, creatinine was the strongest predictor, which is clinically consistent since it is associated with renal function and the diagnosis of CKD. Gender and body mass index (BMI) also demonstrated significant predictive importance which indicates that both biological as well as anthropometric characteristics contribute to the prediction of the disease. Age, as expected, had a moderate influence reflecting its generic risk factor status in chronic diseases.

On the contrary, biochemical parameters like triglycerides (TGs) and cholesterol, as well as comorbidities such as hypertension and myocardial infarction (MI), were relatively less important for the model's decisions. Of note, a marker of long-term glucose control, HbA1c%, played a minimal role which suggests that, within this diabetic population, renal function may be more closely associated with kidney-specific biomarkers rather than a history of glycemic control.

In any case, this is a unique case of feature importance which reaffirms the hypothesis that renal biomarkers are vital in predicting CKD and corroborates the clinical credibility of the internal decision logic employed by the Random Forest model. It further aids in the development of tailored guidelines for screening and proactive intervention in diabetes-dominated high-risk groups.

## 4.6 Validation of Machine Learning Models

The following table shows the validation of Machine learning model features with statistical features.

*Table 4.9 Comparative validation of Features*

| Feature | Statistical Importance (CVD) | ML Importance (CVD) | Statistical Importance (CKD) | ML Importance (CKD) |
|---|---|---|---|---|
| **Creatinine** | Very High | High | High | Highest |
| **HbA1c** | Moderate | Moderate | Moderate | Moderate |
| **BMI** | Moderate | Low | High | High |
| **Cholesterol** | High | High | Low | Low |
| **Triglycerides (TG)** | High | Highest | Low | Minor |
| **Hypertension** | Moderate | Moderate | Low | Low |
| **History of MI** | High | High | Low | Least |
| **History of Stroke** | High | Low | Low | Low |

The table 4.9 shows concordance between the significant features identified through statistical analysis and those flagged as important by the machine learning model provides compelling confirmation of the model's interpretability accuracy and its clinical relevance. Important features like creatinine, cholesterol, and HbA1c showed high importance in both the statistical assessment and the Random Forest model, confirming their pivotal role in predicting cardiovascular and renal complications in diabetic patients. This alignment strengthens the hypothesis that the model is not functioning based on chance correlations, but rather on fundamental biologically and clinically meaningful variables. Some discrepancies also stand out, like the history of stroke or hypertension's importance in predicting CKD; the machine learning model deemphasizes these variables, demonstrating its capacity to neutralize factors of little relevance to the condition being evaluated. These results demonstrate the advantage of using statistical validation together with machine learning to ensure that healthcare models that are predictive in nature are both effective and reliable.

## 4.7 Limitations

This study has a scope of design and statistical evaluation, yet is limited in important areas. First, the available dataset is based on a cross-sectional model which limits assumptions regarding causality as well as the ability to evaluate progression of the disease over time. Second, SMOTE techniques for class imbalance in machine learning were implemented; however, the created samples are unlikely to represent the vast complexity and heterogeneity of real-world patient cases. In addition, some important perplexing variables such as the duration of the diabetes, adherence to prescribed medication, and lifestyle (diet, physical activity) were excluded and could have modified the associations under consideration without any relevance to the absence of these factors. Finally, the building and training of the machine-learning models on these datasets undergoes simulations and afterward randomized tests, which in turn suggests that many other models undergone systems validation that enables them to generalized freely after being tested to other populations.

## 4.8 Conclusion and Future Recommendation

This research study highlighted that Random Forest was the most successful model in predicting Chronic Kidney Disease (CKD) and Cardiovascular Disease (CVD) in patients with diabetes. Its performance exceeded other models due to its accuracy and ability to capture intricate patterns inherent in the data. While Logistic Regression offered ease of interpretation, its lack of accuracy and precision, along with its constant inaccuracy to identify important non-linear relationships, rendered it impractical. The study further demonstrated that crucial medical variables creatinine levels, BMI, cholesterol, triglycerides, HbA1c, and relevant cardiac histories significantly contributed to accurate early disease prediction.

The important aspect of this study is centered on diabetic patients with the integration of machine learning model validation to these features, thus making the derived predictions not only accurate but also possessing clinical relevance. Such provided concentrated data could enhance prompt intervention among a population vulnerable to both CKD and CVD. In constrained-resource regions, such models may assist healthcare providers in early detection of high-risk patients, minimizing unnecessary tests, and elevating patient care. This work expands the use of intelligent systems in medicine, adding to the body of explainable AI, while simultaneously paving the way for developing reliable and efficient claims tools in the future.

Moreover, in order to widen the boundaries of clinical practice with machine learning, future work should focus on obtaining external validation of the models to test them against varied datasets. These models also need to be adjusted to improve description to be provided by explainable AI like SHAP and LIME fundamentally it would improve trust and the transparency for clinicians. The addition of longitudinal and time-based data, such as laboratory data trends over time and treatment histories, could increase the relevance of these tools in practice and their predictive power. Eventually, these models would need to be integrated into the clinical decision support systems to enable proactively identifying patients at high risk of CKD and CVD to provide timely targeted interventions in the management of diabetes.

This research illustrates the synergistic relationship between statistical evaluation and machine learning modeling in the assessment and management of health risks in diabetic individuals. Collectively, they strengthen the foundation for more tailored and effective healthcare interventions.

# REFERENCES

[1] N. A. ElSayed *et al.*, "2. Classification and Diagnosis of Diabetes: Standards of Care in Diabetes-2023," (in eng), *Diabetes Care,* vol. 46, no. Suppl 1, pp. S19-s40, Jan 1 2023, doi: 10.2337/dc23-S002.

[2] "Global, regional, and national burden of diabetes from 1990 to 2021, with projections of prevalence to 2050: a systematic analysis for the Global Burden of Disease Study 2021," (in eng), *Lancet,* vol. 402, no. 10397, pp. 203-234, Jul 15 2023, doi: 10.1016/s0140-6736(23)01301-6.

[3] M. Afkarian *et al.*, "Clinical Manifestations of Kidney Disease Among US Adults With Diabetes, 1988-2014," (in eng), *Jama,* vol. 316, no. 6, pp. 602-10, Aug 9 2016, doi: 10.1001/jama.2016.10924.

[4] J. P. Anderson *et al.*, "Reverse Engineering and Evaluation of Prediction Models for Progression to Type 2 Diabetes:An Application of Machine Learning Using Electronic Health Records," *Journal of Diabetes Science and Technology,* vol. 10, no. 1, pp. 6-18, 2016, doi: 10.1177/1932296815620200.

[5] H. Y. Lu *et al.*, "Digital Health and Machine Learning Technologies for Blood Glucose Monitoring and Management of Gestational Diabetes," *IEEE Reviews in Biomedical Engineering,* vol. 17, pp. 98-117, 2024, doi: 10.1109/RBME.2023.3242261.

[6] V. Jaiswal, A. Negi, and T. Pal, "A review on current advances in machine learning based diabetes prediction," (in eng), *Prim Care Diabetes,* vol. 15, no. 3, pp. 435-443, Jun 2021, doi: 10.1016/j.pcd.2021.02.005.

[7] H. Kwiendacz *et al.*, "Predicting major adverse cardiac events in diabetes and chronic kidney disease: a machine learning study from the Silesia Diabetes-Heart Project," (in eng), *Cardiovasc Diabetol,* vol. 24, no. 1, p. 76, Feb 15 2025, doi: 10.1186/s12933-025-02615-w.

[8] "World Health Organization. Diabetes Key Facts." https://www.who.int/news-room/fact-sheets/detail/diabetes (accessed.

[9] " Report of the fifth meeting of the WHO Technical Advisory Group on Diabetes: hybrid meeting ", Geneva, 7–8 June 2023 2023. [Online]. Available:

https://iris.who.int/bitstream/handle/10665/374877/9789240084810-eng.pdf?sequence=1

[10] M. T. K, S. B. Patil, M. S. B. Ab Wahab, and S. B. Javali, "Screening of Random Blood Sugar in Women: A Critical View," *Journal of Advances in Medicine and Medical Research,* vol. 20, no. 11, pp. 1-5, 04/13 2017, doi: 10.9734/BJMMR/2017/31749.

[11] D. Simmons *et al.*, "Treatment of Gestational Diabetes Mellitus Diagnosed Early in Pregnancy," (in eng), *N Engl J Med,* vol. 388, no. 23, pp. 2132-2144, Jun 8 2023, doi: 10.1056/NEJMoa2214956.

[12] M. Maleki, A. Payandeh, M. Seraji, M. Taherkhani, and A. Bolouri, "International Journal of Public Health Science (IJPHS)," 2023.

[13] C. Krittanawong *et al.*, "Machine learning and deep learning to predict mortality in patients with spontaneous coronary artery dissection," (in eng), *Sci Rep,* vol. 11, no. 1, p. 8992, Apr 26 2021, doi: 10.1038/s41598-021-88172-0.

[14] V. Gulshan *et al.*, "Development and Validation of a Deep Learning Algorithm for Detection of Diabetic Retinopathy in Retinal Fundus Photographs," (in eng), *Jama,* vol. 316, no. 22, pp. 2402-2410, Dec 13 2016, doi: 10.1001/jama.2016.17216.

[15] Y. L. Chiu, M. J. Jhou, T. S. Lee, C. J. Lu, and M. S. Chen, "Health Data-Driven Machine Learning Algorithms Applied to Risk Indicators Assessment for Chronic Kidney Disease," (in eng), *Risk Manag Healthc Policy,* vol. 14, pp. 4401-4412, 2021, doi: 10.2147/rmhp.S319405.

[16] C. Li *et al.*, "Prevalence of painful diabetic peripheral neuropathy in type 2 diabetes mellitus and diabetic peripheral neuropathy: A nationwide cross-sectional study in mainland China," (in eng), *Diabetes Res Clin Pract,* vol. 198, p. 110602, Apr 2023, doi: 10.1016/j.diabres.2023.110602.

[17] D. A. Debal and T. M. Sitote, "Chronic kidney disease prediction using machine learning techniques," *Journal of Big Data,* vol. 9, no. 1, p. 109, 2022/11/20 2022, doi: 10.1186/s40537-022-00657-5.

[18] H. K. Altaf Ali Mangi, Awais Ahmed Junno, Rafiq Ahmed Nindwani, Muhammad Yousuf, Muhammad Shehanshah. , "Knowledge, Attitude and Practice about diabetes among diabetic patients in Sindh, Pakistan.," *RMJ. ,* vol. 43, no. 1, pp. 4-7, December 04, 2023 (2018).

# REFERENCES

[19] K. S. Khokhar GY, Jesrani VK, Gabol IA, Sethar MA., "Prevalence of diabetes in young adults between 18-35 years of the population of Sukkur, Sindh, Pakistan.," *RMJ.,* vol. 48, no. 2, pp. 331-333, December 04, 2023 2023. [Online]. Available: https://rmj.org.pk/fulltext/27-1655485993.pdf?1718812801.

[20] K. Abdul Basit, A. Fawwad, N. Mustafa, T. Davey, B. Tahir, and A. Basit, "Changes in the prevalence of diabetes, prediabetes and associated risk factors in rural Baluchistan; a secondary analysis from repeated surveys (2002-2017)," (in eng), *PLoS One,* vol. 18, no. 4, p. e0284441, 2023, doi: 10.1371/journal.pone.0284441.

[21] A. Khunte *et al.*, "Artificial Intelligence-Based Automated Interpretation of Images of Electrocardiograms: Development and Multinational Validation of ECG-GPT," *medRxiv,* p. 2024.02.17.24302976, 2025, doi: 10.1101/2024.02.17.24302976.

[22] T. T. Aniley, L. K. Debusho, Z. M. Nigusie, W. K. Yimer, and B. B. Yimer, "A semi-parametric mixed models for longitudinally measured fasting blood sugar level of adult diabetic patients," *BMC Medical Research Methodology,* vol. 19, no. 1, p. 13, 2019/01/10 2019, doi: 10.1186/s12874-018-0648-x.

[23] H. A. Al-Alshaikh *et al.*, "Comprehensive evaluation and performance analysis of machine learning in heart disease prediction," *Scientific Reports,* vol. 14, no. 1, p. 7819, 2024/04/03 2024, doi: 10.1038/s41598-024-58489-7.

[24] I. Collein, "The Effect of 4 Pillars of Health Education on Blood Sugar Levels in Type 2 Diabetes Mellitus," *Poltekita : Jurnal Ilmu Kesehatan,* vol. 17, no. 3, pp. 529-534, 10/26 2023, doi: 10.33860/jik.v17i3.2290.

[25] A. V. Khadilkar *et al.*, "Random Blood Glucose Concentrations and their Association with Body Mass Index in Indian School Children," *Indian Journal of Endocrinology and Metabolism,* vol. 23, no. 5, pp. 529-535, 2019, doi: 10.4103/ijem.IJEM_536_19.

[26] M. H. Abernethy, C. Andre, D. W. Beaven, H. W. Taylor, and G. Welsh, "A random blood sugar diabetes detection survey," (in eng), *N Z Med J,* vol. 86, no. 593, pp. 123-126, 1977/08// 1977.

[27] N. C. Fichadiya, A. M. Kadri, and B. B. Dave, "Evaluation of Indian Diabetes Risk Score and Random Blood Sugar Testing for Opportunistic Screening of

Type 2 Diabetes Patients at a District Hospital of Gujarat," (in eng), *Indian J Community Med,* vol. 47, no. 4, pp. 517-521, 2022 Oct-Dec 2022, doi: 10.4103/ijcm.ijcm_1390_21.

[28] U. Md Kafil *et al.*, "MACHINE LEARNING-BASED EARLY DETECTION OF KIDNEY DISEASE: A COMPARATIVE STUDY OF PREDICTION MODELS AND PERFORMANCE EVALUATION," *International Journal of Medical Science and Public Health Research,* vol. 5, no. 12, pp. 58-75, 12/15 2024, doi: 10.37547/ijmsphr/Volume05Issue12-05.

[29] M. A. Islam, M. Z. H. Majumder, and M. A. Hussein, "Chronic kidney disease prediction based on machine learning algorithms," *Journal of Pathology Informatics,* vol. 14, p. 100189, 2023/01/01/ 2023, doi: https://doi.org/10.1016/j.jpi.2023.100189.

[30] X. Li *et al.*, "Machine learning algorithm for predict the in-hospital mortality in critically ill patients with congestive heart failure combined with chronic kidney disease," *Renal Failure,* vol. 46, no. 1, p. 2315298, 2024/12/31 2024, doi: 10.1080/0886022X.2024.2315298.

[31] A. Rajkomar *et al.*, "Scalable and accurate deep learning with electronic health records," *npj Digital Medicine,* vol. 1, no. 1, p. 18, 2018/05/08 2018, doi: 10.1038/s41746-018-0029-1.

[32] T. R. Einarson, A. Acs, C. Ludwig, and U. H. Panton, "Prevalence of cardiovascular disease in type 2 diabetes: a systematic literature review of scientific evidence from across the world in 2007-2017," (in eng), *Cardiovasc Diabetol,* vol. 17, no. 1, p. 83, Jun 8 2018, doi: 10.1186/s12933-018-0728-6.

[33] G. N. Ahmad, H. Fatima, S. Ullah, A. S. Saidi, and Imdadullah, "Efficient Medical Diagnosis of Human Heart Diseases Using Machine Learning Techniques With and Without GridSearchCV," *IEEE Access,* vol. 10, pp. 80151-80173, 2022, doi: 10.1109/ACCESS.2022.3165792.

[34] A. Esteva *et al.*, "A guide to deep learning in healthcare," *Nature Medicine,* vol. 25, no. 1, pp. 24-29, 2019/01/01 2019, doi: 10.1038/s41591-018-0316-z.

[35] T. A. Assegie, P. K. Rangarajan, N. K. Kumar, and D. Vigneswari, "An empirical study on machine learning algorithms for heart disease prediction," *2022,* decision tree; heart disease prediction; random forest; recursive feature

elimination; support vector machine; vol. 11, no. 3, p. 8, 2022-09-01 2022, doi: 10.11591/ijai.v11.i3.pp1066-1073.

[36] Q. Li, H. Lv, Y. Chen, J. Shen, J. Shi, and C. Zhou, "Development and Validation of a Machine Learning Predictive Model for Cardiac Surgery-Associated Acute Kidney Injury," (in eng), *J Clin Med,* vol. 12, no. 3, Feb 1 2023, doi: 10.3390/jcm12031166.

[37] Z. Xiao *et al.*, "Emerging early diagnostic methods for acute kidney injury," (in eng), *Theranostics,* vol. 12, no. 6, pp. 2963-2986, 2022, doi: 10.7150/thno.71064.

[38] I. Kavakiotis, O. Tsave, A. Salifoglou, N. Maglaveras, I. Vlahavas, and I. Chouvarda, "Machine Learning and Data Mining Methods in Diabetes Research," (in eng), *Comput Struct Biotechnol J,* vol. 15, pp. 104-116, 2017, doi: 10.1016/j.csbj.2016.12.005.

[39] T. F. Shoukat F, Shahid S, Ahmad A, Gilani SA. RMJ. (2019), [cited ]; 44(1): 28-31., " Effects of diabetes associated complications on quality of life in patients with type 2 diabetes.," *RMJ.,* vol. 44, no. 1, pp. 28-31, December 05, 2023 2019. [Online]. Available: https://rmj.org.pk/fulltext/27-1521026927.pdf?1718813128.

[40] M. G. Gobena and M. Z. Kassie, "Determinants of blood sugar level among type I diabetic patients in Debre Tabor General Hospital, Ethiopia: a longitudinal study," *Scientific Reports,* vol. 12, no. 1, p. 9035, 2022/05/31 2022, doi: 10.1038/s41598-022-12891-1.

[41] S. Vollmer *et al.*, "Machine learning and artificial intelligence research for patient benefit: 20 critical questions on transparency, replicability, ethics, and effectiveness," (in eng), *Bmj,* vol. 368, p. l6927, Mar 20 2020, doi: 10.1136/bmj.l6927.

[42] M. E. A. Mohammed, S. Alshahrani, G. Zaman, M. Alelyani, I. Hadadi, and M. Musa, "Lipid profile, random blood glucose and carotid arteries thickness in human male subjects with different ages and body mass indexes," *The Aging Male,* vol. 23, no. 5, pp. 1409-1415, 2020/12/04 2020, doi: 10.1080/13685538.2020.1773424.

[43] N. W. Suniyadewi and G. N. I. Pinatih, "Correlation between intakes of carbohydrates, protein, and fat with random blood sugar levels in menopausal

women," *Frontiers of Nursing,* vol. 6, no. 1, pp. 77-80, 2019, doi: doi:10.1515/fon-2018-0041.

[44] N. Razavian, S. Blecker, A. M. Schmidt, A. Smith-McLallen, S. Nigam, and D. Sontag, "Population-Level Prediction of Type 2 Diabetes From Claims Data and Analysis of Risk Factors," *Big Data,* vol. 3, no. 4, pp. 277-287, 2015, doi: 10.1089/big.2015.0020.

[45] S. Asif *et al.*, "Advancements and Prospects of Machine Learning in Medical Diagnostics: Unveiling the Future of Diagnostic Precision," *Archives of Computational Methods in Engineering,* vol. 32, no. 2, pp. 853-883, 2025/03/01 2025, doi: 10.1007/s11831-024-10148-w.

[46] X.-h. Li, X.-l. Yang, B.-b. Dong, and Q. Liu, "Predicting 28-day all-cause mortality in patients admitted to intensive care units with pre-existing chronic heart failure using the stress hyperglycemia ratio: a machine learning-driven retrospective cohort analysis," *Cardiovascular Diabetology,* vol. 24, no. 1, p. 10, 2025/01/08 2025, doi: 10.1186/s12933-025-02577-z.

[47] M. Khalid Hussain, S. Wani, and A. Abubakar, "Examining Mortality Risk Prediction Using Machine Learning in Heart Failure Patients," *International Journal on Perceptive and Cognitive Computing,* vol. 11, no. 1, pp. 81-87, 01/30 2025, doi: 10.31436/ijpcc.v11i1.561.

[48] R. Tr, U. K. Lilhore, P. M, S. Simaiya, A. Kaur, and M. Hamdi, "PREDICTIVE ANALYSIS OF HEART DISEASES WITH MACHINE LEARNING APPROACHES," *Malaysian Journal of Computer Science,* pp. 132-148, 03/31 2022, doi: 10.22452/mjcs.sp2022no1.10.

[49] R. Khera *et al.*, "Use of Machine Learning Models to Predict Death After Acute Myocardial Infarction," (in eng), *JAMA Cardiol,* vol. 6, no. 6, pp. 633-641, Jun 1 2021, doi: 10.1001/jamacardio.2021.0122.

[50] A. Charleonnan, T. Fufaung, T. Niyomwong, W. Chokchueypattanakit, S. Suwannawach, and N. Ninchawee, "Predictive analytics for chronic kidney disease using machine learning techniques," in *2016 Management and Innovation Technology International Conference (MITicon)*, 12-14 Oct. 2016 2016, pp. MIT-80-MIT-83, doi: 10.1109/MITICON.2016.8025242.

[51] V. R. Kutty, C. R. Soman, A. Joseph, K. V. Kumar, and R. Pisharody, "Random Capillary Blood Sugar and Coronary Risk Factors in a South Kerala

Population," *Journal of Cardiovascular Risk,* vol. 9, no. 6, pp. 361-367, 2002, doi: 10.1177/174182670200900610.

[52] S. T. Marwah and T. R. Marwah, "A Study on Correlation of Random Blood Sugar Levels with Cardiovascular Outcome in Patients of Myocardial Infarction at a Tertiary Care Centre," (in eng), *J Assoc Physicians India,* vol. 70, no. 4, pp. 11-12, 2022/04// 2022. [Online]. Available: http://europepmc.org/abstract/MED/35443403.

[53] H. G. Mendhe, H. Narni, S. C. P, and S. M, "Obesity indices comparison and its correlation with random blood sugar and blood pressure in adults in rural field practice area of a medical college," *International Journal Of Community Medicine And Public Health,* vol. 3, no. 9, pp. 2555-2560, 12/24 2016, doi: 10.18203/2394-6040.ijcmph20163071.

[54] A. Kumar, N. Kumar, and A. Kumar, "Evaluation of Random Blood Sugar in Chronic Liver Disease Patient of Bihar," (in English), *European Journal of Molecular and Clinical Medicine,* Report vol. 9, p. 1352+, 2022 Wntr // 2022. [Online]. Available: https://link.gale.com/apps/doc/A698246007/HRCA?u=anon~894ce3a8&sid=googleScholar&xid=9b404e7e.

[55] J. J. Niveda and R. Y. Rajkumar, "Comparative Analysis of Machine Learning Algorithms for Early Detection of Chronic Kidney Disease: Performance Evaluation and Insights," in *2024 Third International Conference on Smart Technologies and Systems for Next Generation Computing (ICSTSN)*, 18-19 July 2024 2024, pp. 1-6, doi: 10.1109/ICSTSN61422.2024.10671021.

[56] A. S. Tan, L. S. Yong, S. Wan, and M. L. Wong, "Patient education in the management of diabetes mellitus," (in eng), *Singapore Med J,* vol. 38, no. 4, pp. 156-160, 1997/04// 1997. [Online]. Available: http://europepmc.org/abstract/MED/9269394.

[57] N. Gautam *et al.*, "Machine Learning in Cardiovascular Risk Prediction and Precision Preventive Approaches," (in eng), *Curr Atheroscler Rep,* vol. 25, no. 12, pp. 1069-1081, Dec 2023, doi: 10.1007/s11883-023-01174-3.

[58] A. J. Steele, S. C. Denaxas, A. D. Shah, H. Hemingway, and N. M. Luscombe, "Machine learning models in electronic health records can outperform conventional survival models for predicting patient mortality in coronary artery

disease," *PLOS ONE,* vol. 13, no. 8, p. e0202344, 2018, doi: 10.1371/journal.pone.0202344.

[59] S. F. Weng, J. Reps, J. Kai, J. M. Garibaldi, and N. Qureshi, "Can machine-learning improve cardiovascular risk prediction using routine clinical data?," *PLOS ONE,* vol. 12, no. 4, p. e0174944, 2017, doi: 10.1371/journal.pone.0174944.

[60] S.-Y. Cho *et al.*, "Pre-existing and machine learning-based models for cardiovascular risk prediction," *Scientific Reports,* vol. 11, no. 1, p. 8886, 2021/04/26 2021, doi: 10.1038/s41598-021-88257-w.

[61] K. W. Johnson *et al.*, "Artificial Intelligence in Cardiology," *JACC,* vol. 71, no. 23, pp. 2668-2679, 2018, doi: doi:10.1016/j.jacc.2018.03.521.

[62] B. Khan, R. Naseem, F. Muhammad, G. Abbas, and S. Kim, "An Empirical Evaluation of Machine Learning Techniques for Chronic Kidney Disease Prophecy," *IEEE Access,* vol. 8, pp. 55012-55022, 2020, doi: 10.1109/ACCESS.2020.2981689.

[63] C. Thongprayoon *et al.*, "Promises of Big Data and Artificial Intelligence in Nephrology and Transplantation," *Journal of Clinical Medicine,* vol. 9, p. 1107, 04/13 2020, doi: 10.3390/jcm9041107.

[64] J. Theis, W. L. Galanter, A. D. Boyd, and H. Darabi, "Improving the In-Hospital Mortality Prediction of Diabetes ICU Patients Using a Process Mining/Deep Learning Architecture," (in eng), *IEEE J Biomed Health Inform,* vol. 26, no. 1, pp. 388-399, Jan 2022, doi: 10.1109/jbhi.2021.3092969.

[65] M. Afshar *et al.*, "Natural language processing and machine learning to identify alcohol misuse from the electronic health record in trauma patients: development and internal validation," (in eng), *J Am Med Inform Assoc,* vol. 26, no. 3, pp. 254-261, Mar 1 2019, doi: 10.1093/jamia/ocy166.

[66] Q. Zou, K. Qu, Y. Luo, D. Yin, Y. Ju, and H. Tang, "Predicting Diabetes Mellitus With Machine Learning Techniques," (in English), *Frontiers in Genetics,* Original Research vol. Volume 9 - 2018, 2018-November-06 2018, doi: 10.3389/fgene.2018.00515.

[67] M. Alghamdi, M. Al-Mallah, S. Keteyian, C. Brawner, J. Ehrman, and S. Sakr, "Predicting diabetes mellitus using SMOTE and ensemble machine learning

approach: The Henry Ford ExercIse Testing (FIT) project," (in eng), *PLoS One,* vol. 12, no. 7, p. e0179805, 2017, doi: 10.1371/journal.pone.0179805.

[68]     P. B. Jensen, L. J. Jensen, and S. Brunak, "Mining electronic health records: towards better research applications and clinical care," *Nature Reviews Genetics,* vol. 13, no. 6, pp. 395-405, 2012/06/01 2012, doi: 10.1038/nrg3208.

[69]     J. Yoon, W. R. Zame, A. Banerjee, M. Cadeiras, A. M. Alaa, and M. van der Schaar, "Personalized survival predictions via Trees of Predictors: An application to cardiac transplantation," *PLOS ONE,* vol. 13, no. 3, p. e0194985, 2018, doi: 10.1371/journal.pone.0194985.

[70]     A. Y. Hannun *et al.*, "Cardiologist-level arrhythmia detection and classification in ambulatory electrocardiograms using a deep neural network," (in eng), *Nat Med,* vol. 25, no. 1, pp. 65-69, Jan 2019, doi: 10.1038/s41591-018-0268-3.

[71]     R. Miotto, L. Li, B. A. Kidd, and J. T. Dudley, "Deep Patient: An Unsupervised Representation to Predict the Future of Patients from the Electronic Health Records," *Scientific Reports,* vol. 6, no. 1, p. 26094, 2016/05/17 2016, doi: 10.1038/srep26094.

[72]     M. T. Ribeiro, S. Singh, and C. Guestrin, ""Why Should I Trust You?": Explaining the Predictions of Any Classifier," presented at the Proceedings of the 22nd ACM SIGKDD International Conference on Knowledge Discovery and Data Mining, San Francisco, California, USA, 2016. [Online]. Available: https://doi.org/10.1145/2939672.2939778.

[73]     Z. I. Attia *et al.*, "An artificial intelligence-enabled ECG algorithm for the identification of patients with atrial fibrillation during sinus rhythm: a retrospective analysis of outcome prediction," (in eng), *Lancet,* vol. 394, no. 10201, pp. 861-867, Sep 7 2019, doi: 10.1016/s0140-6736(19)31721-0.

[74]     A. Goldstein, K. Adam, B. Justin, and E. and Pitkin, "Peeking Inside the Black Box: Visualizing Statistical Learning With Plots of Individual Conditional Expectation," *Journal of Computational and Graphical Statistics,* vol. 24, no. 1, pp. 44-65, 2015/01/02 2015, doi: 10.1080/10618600.2014.907095.

[75]     L. Breiman, "Random Forests," *Machine Learning,* vol. 45, no. 1, pp. 5-32, 2001/10/01 2001, doi: 10.1023/A:1010933404324.

[76] E. J. Topol, "High-performance medicine: the convergence of human and artificial intelligence," (in eng), *Nat Med,* vol. 25, no. 1, pp. 44-56, Jan 2019, doi: 10.1038/s41591-018-0300-7.

[77] B. Draznin *et al.*, "2. Classification and Diagnosis of Diabetes: Standards of Medical Care in Diabetes-2022," *Diabetes care,* vol. 45, pp. S17-S38, 01/01 2022, doi: 10.2337/dc22-S002.

[78] G. H. Tison *et al.*, "Passive Detection of Atrial Fibrillation Using a Commercially Available Smartwatch," (in eng), *JAMA Cardiol,* vol. 3, no. 5, pp. 409-416, May 1 2018, doi: 10.1001/jamacardio.2018.0136.

[79] C. Cortes and V. Vapnik, "Support-vector networks," *Machine Learning,* vol. 20, no. 3, pp. 273-297, 1995/09/01 1995, doi: 10.1007/BF00994018.

[80] M. A. Asif *et al.*, "Performance Evaluation and Comparative Analysis of Different Machine Learning Algorithms in Predicting Cardiovascular Disease," *Engineering Letters,* vol. 29, pp. 731-741, 05/17 2021.